%% file: 0main.tex
\documentclass[final,authoryear,12pt,fleqn]{elsarticle}
\usepackage{graphicx} % Required for inserting images
\usepackage[utf8]{inputenc}
\usepackage{bm}
\usepackage{natbib}
\usepackage[hidelinks]{hyperref}
\usepackage{xcolor}
\usepackage{enumitem}
\usepackage{amsmath}
\DeclareMathOperator*{\argmax}{arg\,max}

\usepackage{amsfonts}
\usepackage{rotating}
\usepackage[ruled,vlined]{algorithm2e}
\usepackage{caption}
\usepackage{subcaption}
\usepackage[a4paper,margin=2.5cm]{geometry}
\usepackage{float}
\usepackage{makeidx}
\usepackage{setspace}
\usepackage{booktabs}

\makeatletter
\def\ps@pprintTitle{%
 \let\@oddhead\@empty
 \let\@evenhead\@empty
 \def\@oddfoot{}%
 \let\@evenfoot\@oddfoot}

\begin{document}
\begin{frontmatter}

\title{\huge{Forecasting Oil Consumption: The Statistical Review of World Energy Meets
Machine Learning}}
\author[1]{Jan Ditzen}
\ead{jan.ditzen@unibz.it}
\author[2]{Erkal Ersoy}
\ead{e.ersoy@hw.ac.uk}
\author[2]{Haoyang Li}
\ead{hl2030@hw.ac.uk}
\author[1,3]{Francesco Ravazzolo}
\ead{francesco.ravazzolo@unibz.it}
\affiliation[1]{Free University of Bozen-Bolzano}
\affiliation[2]{Heriot-Watt University}
\affiliation[3]{BI Norwegian Business School}
\date{November 2025}

\tnotetext[acknowledgements]{Acknowledgements: For comments and discussions, we are grateful to the participants of the “Econometric and Macro-Financial
Models of Climate Change and Energy Markets” workshop at the University of Bolzano. Jan Ditzen acknowledges financial support from Italian Ministry MIUR under the PRIN-PNRR “Consumer behavior, digital platforms and spatial spillover effects in a dual electricity market (SPEX)” Codice progetto: P20227RKF9 - CUP: I53D23006190001.}

\begin{abstract}
    This paper studies whether a small set of dominant countries can account for most of the dynamics of regional oil demand and improve forecasting performance. We focus on dominant drivers within the OECD and a broad GVAR sample covering over 90\% of world GDP. Our approach identifies dominant drivers from a high-dimensional concentration matrix estimated row by row using two complementary variable-selection methods, LASSO and the one-covariate-at-a-time multiple testing (OCMT) procedure. Dominant countries are selected by ordering the columns of the concentration matrix by their norms and applying a criterion based on consecutive norm ratios, combined with economically motivated restrictions to rule out pseudo-dominance. The United States emerges as a global dominant driver, while France and Japan act as robust regional hubs representing European and Asian components, respectively. Including these dominant drivers as regressors for all countries yields statistically significant forecast gains over autoregressive benchmarks and country-specific LASSO models, particularly during periods of heightened global volatility. The proposed framework is flexible and can be applied to other macroeconomic and energy variables with network structure or spatial dependence.
\end{abstract}

\begin{keyword}
Oil demand; LASSO; OCMT; High-dimensional; Dominant drivers\\

\textit{JEL classification:} C13; C33
\end{keyword}
\end{frontmatter}

\clearpage
\doublespacing

\section{Introduction} \label{sec: intro}
\input{1Intro}

\section{Methodology} \label{sec: Method}
\input{2Methodology}

\section{Data and models} \label{sec: data model}
\input{3DataModel}

\section{Results} \label{sec: results}
\input{4Results}

\section{Summary} \label{sec: conclusion}
\input{5Summary}

\clearpage

%%TC:ignore
\bibliographystyle{chicago}
%\addcontentsline{toc}{chapter}{Bibliography}
%\pagestyle{ref}
\bibliography{ref.bib}
%%TC:endignore

\input{6Appendix}

\end{document}

%% file: 1Intro.tex
Since the 1970s, world oil consumption has more than doubled with the growth overwhelmingly driven by developing economies. Historical data from the Energy Institute's 2025 Statistical Review of World Energy show that global oil consumption rose from around 45.7 million barrels per day (mb/d) in 1970 to 101.4 mb/d in 2024, with 10.2 mb/d growth from OECD countries and 45.6 mb/d from non-OECD economies, accounting for 81.8\% of the total growth. While the consumption of fossil fuels has led to challenges such as global warming, air pollution, and health risks, oil remains the largest source of energy and accounted for approximately 34\% of total final energy demand in 2024. OECD countries took the lead in investing in low-carbon technology, which improves energy efficiency and helps slow down overall oil consumption growth \citep{paramati2022role}. 

Many OECD members already have relatively stable, or even flat, oil demand growth. The rich literature on the subject has identified income and fuel prices as the main determinants of oil consumption \citep{ozcan2015determinants, gao2021income}, with recent evidence using long quarterly panels for 21 OECD countries confirming that both income and price elasticities are small but statistically significant and time-varying in determining demand for oil \citep{helmi2024time}. These studies treat OECD as a homogeneous block, implicitly assuming an equal contribution to regional demand dynamics from all countries. However, data show that OECD oil demand is highly concentrated in a few large economies, with the United states accounting for 19 mb/d in 2024, followed by Japan, South Korea, Canada, Germany, and France. \cite{zhu2021cross} tried to model cross-border risk spillovers in the global oil system and found that a small set of countries plays a large role in transmitting oil shocks through trade and financial channels.

Inspired by the idea that idiosyncratic firm-level fluctuations can explain an important part of aggregate shocks \citep{gabaix2011granular}, this paper shifts focus from elasticities of income and price to dominant countries within OECD that may provide additional gains in predicting regional oil demand. If several economies' consumption dynamics can explain most of the variation in aggregate OECD demand and transmit oil-related global shocks to a large share of countries, including them in the model can not only improve prediction accuracy but also provide more robust predictions at the time of unexpected shocks.

While doing this, we include more than just oil consumption in the list of potential dominant drivers. Country-specific macroeconomic and energy variables can also serve as dominant drivers. Consequently, the set of candidate dominant drivers can exceed the number of periods, rendering traditional models invalid. A machine learning algorithm and a novel statistical approach are employed to identify dominant drivers affecting oil demand of OECD members in this paper. Both approaches are also applied to GVAR countries \citep{mohaddes2024compilation}, which cover more than 90\% of world GDP. Dominant drivers share similarities with common factors that affect all countries. Therefore, effects of dominant drivers can be absorbed when common factors are controlled for. We find that the United States acts a global dominant driver for OECD countries, while France and Japan appear to be dominant drivers representing European factor and Asian factor respectively. This finding is robust to different estimation approaches and different samples. Forecasts based on our approach show clear improvement over an AR benchmark. A country-specific LASSO which selects each country's own regressors is also included as competitor, which shows comparable performance in smooth periods but is outperformed by our approach in volatile periods with global shocks. Furthermore, we study the dynamic transmission of shocks to the dominant drivers in a structural VAR and confirm the importance of an Asian hub represented by Japan or Korea to aggregate oil consumption.

Our proposed approach outlined in Algorithm \ref{alg:dominant-drivers} in the Appendix involves two steps. The first step is estimation of the concentration matrix, where a non-zero element indicates a connection between two variables, while a zero element implies no connection after partialling out all other regressors and factors. As such, the entries can be represented by partial correlations and are therefore related to estimated coefficients from regression \citep{sulaimanov2016graph}.

Since the number of candidate dominant drivers is allowed to exceed the number of observations, a high-dimensional variable selection approach is needed. In addition to the rigorous LASSO \citep{Bickel2009a, Belloni2016, Chernozhukov2019, Ahrens2020} and adaptive LASSO \citep{Zou2006, Medeiros2016} adopted by \cite{ditzen2022dominant} to detect dominant drivers for national inflation, we employ a novel statistical method, One Covariate at a Time Multiple Testing (OCMT), proposed by \cite{chudik2018one}. After performing variable selection for each variable, post estimation with selected variables is conducted to estimate coefficients, which are in turn used to construct the concentration matrix \citep{sulaimanov2016graph}. The OCMT estimator is deliberately constructed to have a different selection strategy from LASSO, so that common selected countries across the two methods reflect true structural dominance rather than estimator-specific coincidence.

The next step involves selecting dominant drivers through the concentration matrix. Following \cite{brownlees2021detecting}, we order the columns of the concentration matrix in descending order by column norm then apply a criterion similar to the eigenvalue approach \citep{ahn2013eigenvalue} to identify dominant drivers.

Our work extends the application of dominant drivers to prediction in the context of energy demand. The implementation of two complementary high-dimensional concentration matrix estimators further enhances the reliability of dominant-driver detection in energy applications. We find that a small set of dominant countries--most notably the United States as a global driver and France and Japan as robust regional hubs--captures a large share of oil-demand dynamics, delivers statistically significant forecasting gains over standard benchmarks during periods of heightened global volatility, and plays a central role in the structural transmission of oil-demand shocks to aggregate consumption.

The rest of the paper is structured as follows: Section \ref{sec: Method} describes the steps of concentration matrix estimation and dominant drivers selection in detail. Section \ref{sec: data model} introduces data and models. Section \ref{sec: results} presents our main results, and Section \ref{sec: conclusion} concludes.

%% file: 2Methodology.tex
\subsection{Dominant Drivers Detection} \label{section: DD detection}
Consider a panel data model with $N$ units and $T$ periods:
\begin{equation} \label{eq: panel model}
    x_{it} = \sum_{j \neq i}^N b_{ij} x_{jt} + u_{it}, \quad i,j = 1,2,...,N,\; t = 1,2,...,T,
\end{equation}
where $b_{ij}$ measures the effect of $x_{jt}$ on $x_{it}$.
Or in matrix notation
\begin{equation}
    \bm{x_t = B x_t + u_t},
\end{equation}
where $\bm{x_t} = (x_{1t}, x_{2t}, \ldots, x_{Nt})'$, $\bm{u}_t = (u_{1t}, \ldots, u_{Nt})'$, 
and $\bm{B} = (b_{ij})_{i,j=1}^N$ is the $N \times N$ matrix of pairwise interaction coefficients, with $b_{ii}=0$. Following  \cite{KapetaniosPesaranReese2020} and \cite{ditzen2022dominant}, dominant drivers $j  \in \Gamma(N_d)$ are defined as 
\begin{equation*}
    \lim_{N \rightarrow \infty} \frac{1}{N} \sum_{i=1}^N |b_{ij}| = K > 0,
\end{equation*}
while non-dominant drivers $j \notin \Gamma(N_d)$ are defined as 
\begin{equation*}
    \lim_{N \rightarrow \infty} \sum_{i=1}^N |b_{ij}| = K < \infty.
\end{equation*}
This definition ensures that the impacts of dominant drivers do not diminish as $N \rightarrow \infty$, while the overall influence of non-dominant drivers is bounded. The definition of dominant and non-dominant drivers is corollary to the definition of strong and weak cross-sectional dependence in \cite{chudik2011weak}.  Provided that the number of dominant drivers $N_d$ and the set $\Gamma(N_d)$ are known, the dominant and non-dominant drivers can be defined as in \cite{KapetaniosPesaranReese2020} and \cite{brownlees2021detecting}:
\begin{align}
    x_{it,d} &= u_{it}, \quad i \in \Gamma(N_d) \\
    x_{it,nd} &= \sum_{j \in \Gamma(N_d)} b_{ij}u_{jt} + u_{it}, \quad i \notin \Gamma(N_d) \label{eq: dd def}
\end{align}

In light of this, the identification of dominant drivers lies in the estimation of $N_d$ and $\Gamma(N_d)$, which can be obtained using the error variance \citep{KapetaniosPesaranReese2020} or concentration matrix \citep{brownlees2021detecting}. Under the assumption that $\bm{B}$ is sparse, the non-zero elements in $\bm{B}$ can be estimated using variable selection methods like LASSO and OCMT row by row for all $i$:
\begin{equation} \label{eq: row regression}
    x_{it} = \sum_{j \neq i}^N \hat{b}_{ij} x_{jt} + \hat{\epsilon}_{it}.
\end{equation}
Following \cite{sulaimanov2016graph}, the concentration matrix can be constructed by
\begin{equation} \label{eq: concentration matrix}
    \bm{\hat{\Theta} = \hat{D}(I-\hat{G})\hat{D}},
\end{equation}
where $\bm{\hat{D}}$ is a $N$ by $N$ diagonal matrix with $1/ \hat{\sigma_i}$ on the diagonal, $\hat{\sigma_i}$ is the standard error of the residuals in regression \eqref{eq: row regression} and $\bm{G}$ is a matrix with zeros on the diagonal. This scaling ensures that the concentration matrix is properly normalised by the idiosyncratic variability of each unit. The construction of $\bm{\hat{\Theta}}$ follows the graphical modelling literature, where elements in the concentration matrix represent conditional dependence. Specifically, zeros in $\bm{\hat{\Theta}}$ correspond to pairs of conditionally independent units given all other units. Under the sparsity assumption and the standard stability condition that the spectral radius $\|\bm{G}\| < 1$, $(\bm{I - \hat{G}})$ ensures that the sparsity is carried over to $\bm{\hat{\Theta}}$, and that the associated covariance matrix $\bm{\Sigma}=\bm{\hat{\Theta}}^{-1}$ is positive definite. This is analogous to the role of $(\bm{I} - \rho \bm{W})$ in spatial econometrics, which captures direct dependence, while its inverse aggregates indirect linkages of all orders.
$\bm{\hat{G}}$ deserves further discussion. The individual elements of the concentration matrix $\Theta_{ij}$ can related to the partial correlations as $\rho_{ij}=-\theta_{ij}/\sqrt{\theta_{ii}\theta_{jj}}$ \citep{sulaimanov2016graph,brownlees2021detecting}. Hence, for the non-diagonal elements in $\bm{\Theta}$, represented by $\bm{G}$ we can state that $b_{ij}=g_{ij}$, for $ i\neq j$, and $\bm{B}=\bm{G}$. However, the estimated matrix of coefficients, $\bm{\hat{B}}$ is non-symmetric. To achieve symmetry \cite{Meinshausen2006,sulaimanov2016graph} propose an OR or AND operation. The $i-jth$ entry in the concentration matrix is non-zero under the OR operation if either the $i-jth$ element or the $j-ith$ element is zero; or in the case of the AND operation both elements are non-zero. We follow the less strict OR operation and  arrive at a non-symmetric estimate of the concentration matrix, which we will name \textit{network matrix} henceforth:
\begin{align}
 \bm{\hat{\kappa} = \hat{D}(I-\hat{B})\hat{D}},   
\end{align}

Then, dominant drivers selection can be performed with the estimated concentration matrix. As proposed by \cite{brownlees2021detecting}, the eigenvalue criterion developed by \cite{ahn2013eigenvalue} is applied to the sorted column norms of the concentration matrix in descending order. As an alternative to the sample concentration matrix with $N\leq T$, \cite{brownlees2021detecting} propose to estimate the concentration matrix with a regularized estimator following \cite{Meinshausen2006,Fan2016}. We follow this approach and use our non-symmetric estimate of the concentration matrix $\hat{\bm{\kappa}}$ instead.\footnote{Symmetry for the consistent estimation of the dominant units is not strictly required in \cite{brownlees2021detecting} as long as the concentration matrix is bounded, see their Assumptions 1-3, and the estimated coefficient matrices are estimated consistently. Boundness is met with sufficient standardization, see for example \cite{Pesaran2021a}.} Specifically, the estimated number of dominant drivers is:
\begin{equation}
    \hat{N}_d = \argmax_{s=1,2,\ldots,N-1} \frac{\|\bm{\hat{\kappa}_{(s)}}\|}{\|\bm{\hat{\kappa}_{(s+1)}}\|},
\end{equation}
where $\|\bm{\hat{\kappa}_{(s)}}\|$ is the $s$-th largest column norm of $\bm{\hat{\kappa}}$. Intuitively, the column norm measures the overall strength of a unit's contribution to the panel. Column norms of dominant drivers grow with the number of units $N$, while those of non-dominant drivers remain bounded. Consequently, the ``gap'' between dominant and non-dominant drivers is revealed in the sequence of ordered column norms. The largest drop in the sequential column norm ratio is expected to occur at the transition point between the last dominant driver and the first non-dominant driver. Naturally, selecting $\hat{N}_d$ as the index maximising the ratio provides a data-driven rule to separate dominant drivers from non-dominant drivers. The first $\hat{N}_d$ units are classified as dominant drivers.
In the presence of unobserved common factors, identification of dominant drivers is threatened by confounding. One must partial out the common factors before ranking column norms to distinguish dominant drivers from non-dominant ones. After the adjustment, the loadings $b_{ij}$ on non-dominant drivers must remain sufficiently strong to be identified \citep{brownlees2021detecting}.

However, note that $\bm{\hat{\kappa}}$ has $1/\hat{\sigma}_i^2$ on its diagonal. This creates a risk that a non-dominant unit with very small residual variance $\hat{\sigma}_i^2$ may be falsely classified as dominant due to the inflated column norm by the diagonal element. To address this, define
\begin{equation} \label{eq: diagonal ratio}
    R_i = \frac{|\hat{\kappa}_{(ii)}|}{\|\bm{\hat{\kappa}_{(i)}}\|},
\end{equation}
where $|\hat{\kappa}_{(ii)}|$ is the absolute value of the $i$-th diagonal element and $\|\bm{\hat{\kappa}_{(i)}}\|$ is the norm of the $i$-th column.  The ratio $R_i$ measures how much of a column’s strength is driven by its diagonal element relative to its overall norm. If $R_i$ is close to one, the column’s magnitude is dominated by its own residual variance rather than by spillover effects on other units, making it less credible as a dominant driver. By setting a threshold $R \in [0,1]$, units with $R_i \geq R$ can be excluded from the pool of dominant-driver candidates, thereby filtering out false classifications arising purely from small residual variances.

Another challenge is that the procedure may occasionally classify units with only a few connections as dominant drivers, since the selection criterion does not explicitly account for network connectivity. To mitigate this, we restrict the pool of candidates to those which have more than the median number of links \citep{KapetaniosPesaranReese2020}. This ensures that dominance is not attributed to isolated units but to those with sufficient connectivity to plausibly transmit shocks across the system.

\subsection{Network (Non-Symmetric Concentration) Matrix Estimation}

Consider the panel model in \eqref{eq: panel model}, where the set of $k_i$ true regressors for unit $i$ is nested within a candidate set of dimension $p = N-1 > T$. The network matrix is estimated row by row \citep{Meinshausen2006}. Since the number of candidate dominant drivers may exceed the sample size, we employ two variable selection methods---LASSO and OCMT---to recover the nonzero elements in each row. 

\textbf{LASSO}: This widely used method has been shown to deliver consistent variable selection and post-estimation in both cross-sectional \citep{Bickel2009a} and time-series \citep{Belloni2016,Ahrens2020} settings. The estimator is obtained by solving
\begin{equation} \label{eq: rlasso}
    \hat{\bm{b}}_i 
    = \arg\min_{\bm{b_i} \in \mathbb{R}^{p}} 
    \frac{1}{T} \sum_{t=1}^T \left( x_{it} - \bm{x}_{t}' \bm{b}_i \right)^2 
    + \lambda_i \sum_{j \neq i} \psi_{ij} | b_{ij} |,
\end{equation}
where $b_{ii} = 0$, $\lambda_i$ is a data-driven penalty parameter, and $\psi_{ij}$ are regressor-specific loadings that account for potential heteroskedasticity and correlation among regressors. Equation \eqref{eq: rlasso} corresponds to the rigorous LASSO \citep{Bickel2009a, Belloni2016, Chernozhukov2019, Ahrens2020}, which excludes irrelevant regressors with high probability while retaining the relevant ones. Repeating this estimation for each unit $i$ yields the post-LASSO estimates $\hat{\bm{b}}_i$, which are collected as the $i$th row of the $N \times N$ matrix $\hat{B}$.

\textbf{OCMT}: The One Covariate at a Time Multiple Testing (OCMT) procedure of \citet{chudik2018one} offers an alternative approach to variable selection in high-dimensional settings. Instead of imposing a joint penalty across all candidate regressors, OCMT evaluates each regressor individually. Specifically, for each unit $i$ and candidate regressor $j \neq i$, the auxiliary regression
\begin{equation} \label{eq: ocmt}
    x_{it} = \alpha_{ij} + \delta_{ij} x_{jt} + e_{ijt}, 
    \quad t=1,\dots,T,
\end{equation}
is estimated, and robust $t$-statistics are constructed for each $\delta_{ij}$. The corresponding $p$-values $\{p_{ij}\}_{j \neq i}$ are then adjusted using multiple-testing procedures \citep{bonferroni1936,holm1979simple,benjamini1995controlling,benjamini2001} to control either the family-wise error rate (FWER) or the false discovery rate (FDR) at a pre-specified significance level $\alpha$. Regressors with adjusted $p_{ij} \leq \alpha$ are retained, and the joint model is re-estimated on the selected set.

OCMT delivers consistent variable selection even when the number of potential regressors $p$ grows faster than the sample size $T$, but requires the number of noise variables correlated with true regressors (``pseudo signals'') to rise no faster than $\sqrt{T}$. \citet{Sharifvaghefi2024VarSelectCorrelated} relaxed the assumption by controlling the unobserved common factors and provided theoretical guarantees for this generalised OCMT (GOCMT). Specifically, the regression becomes
\begin{equation} \label{eq: ocmt factors}
        x_{it} = \alpha_{ij} + \bm{\hat{f}_t' \gamma_i} + \delta_{ij} x_{jt} + e_{ijt}, 
    \quad t=1,\dots,T,
\end{equation}
where $\bm{\hat{f}_t}$ is an $m \times 1$ vector of unobserved common factors at period $t$ which can be estimated via principal components when the factors are of interest or controlled by common correlated effects (CCE) \citep{pesaran2006estimation} when recovering the latent common factors is not required. \citet{pesaran2024high} applied the GOCMT procedure to forecast UK inflation. Since estimated unobserved common factors can absorb effects from dominant drivers defined in \eqref{eq: dd def}, we leave the common factors uncontrolled on purpose in our main analysis and report results with common factors controlled via CCE as a robustness check.
Our method is robust to heteroskedasticity and weak dependence, while remaining computationally efficient and transparent, since selection is based on classical hypothesis testing. The selected coefficient vectors $\hat{\bm{b}}_i$ are subsequently collected to form the estimated coefficient matrix $\hat{B}$.

Then, the network matrix $\bm{\hat{\kappa}}$ can be constructed as in \eqref{eq: concentration matrix}. Depending on the data, the dominant drivers can be country-specific regressors or common variables. The procedure is summarised in Algorithm \ref{alg:dominant-drivers}.

%% file: 3DataModel.tex
\subsection{Data}
The core dataset is the \textit{Statistical Review of World Energy 2025} (Energy Institute), which reports country--year series on energy demand and production from 1965 to 2024. We retain the following variables: oil consumption (OC) and production (OP) in thousand barrels per day, coal and natural gas consumption (CoalC, GasC, in exajoules), nuclear and renewables consumption (Nuclear, Renewables, in exajoules), refinery capacity in thousand barrels per day (Refcap), population (Pop), GDP, primary energy intensity (Eintensity), which is the ratio between total primary energy consumption and GDP, and the West Texas Intermediate crude oil price (WTI\_price). Due to data availability, OECD and GVAR countries are considered in this application.

For each year, We define the OECD (GVAR) aggregate oil consumption as the sum of oil consumption across the countries listed in Tables \ref{tab:oecd} and \ref{tab:gvar}:
\begin{equation} \label{eq: agg oil cons}
   Total_{OC,t}^{(g)} = \sum_{i \in \mathcal{I}_g} OC_{it}, \quad g \in \{\text{OECD,\;GVAR}\}  
\end{equation}

where $\mathcal{I}_g$ denotes the set of countries specified in the corresponding table. A $\Delta \log$ transform is applied to all variables. To guard against log zero, variables containing zeros are shifted by a small constant $\varepsilon = 10^{-8}$ prior to logging. We then difference within country and get $\Delta \ln x_{it} = \ln x_{it} - \ln x_{i,t-1}$ for all variables. Any remaining missing values after differencing are set to zero, which implicitly assumes zero growth rates in those missing periods. We also use cross-sectional average variables for each differenced series  to control unobserved common factors \citep{pesaran2006estimation}.
 By design, WTI\_price, Total\_OC, and Total\_OP are excluded from this PC construction as they are common to all countries.
 
The first 70\% of periods are used as the training sample to detect dominant drivers and estimate all models, while the remaining 30\% serve as a hold-out sample for prediction. Since all variables are expressed in log-differences, they are directly comparable across countries and variables, ensuring that the LASSO penalty is not mechanically influenced by scale differences. Additional normalisation is therefore not required, though we implement it to account for potential heteroskedasticity across series.

\subsection{Dominant drivers} \label{sec: DD method}
We identify dominant drivers by estimating the network matrix\footnote{\citet{barigozzi2024fnets} estimates dynamic (lagged) networks and long-run partial-correlation networks via innovation and long-run precision matrices. We focus on contemporaneous dependence because extending our high-dimensional panel setting to lagged network dynamics and long-run covariance estimation would substantially increase modelling complexity and finite sample instability without being necessary for our objective of detecting dominant drivers, especially when $T$ is small.}, as outlined in Section~\ref{section: DD detection}. For each country-specific oil demand growth rate, defined as the difference in logs
\[\Delta \ln OC_{it} = \ln OC_{it} - \ln OC_{it-1},\]
we estimate
\begin{equation}
    \Delta \ln OC_{it} = \alpha_i + \sum_{j \neq i}^N w_{ij} \Delta \ln OC_{jt} + \sum_{j=1}^N \left(\Delta \ln \bm{X_{jt}'} \right) \bm{\beta_{j}} + \bm{f_t' \gamma_i} +u_{it}, 
\end{equation}
where $w_{ij}$ measures the effect of changes in country $j$'s oil demand on country $i$'s oil demand; regressors $\bm{X}$ includes OP, CoalC, GasC, Nuclear, Renewables, Refcap, Pop, GDP, and Eintensity; $\bm{\beta_{i}}$ is the vector of coefficients that captures the effects of changes in $\ln \mathbf{X}_{jt}$ on oil demand growth in country $i$; $\bm{f_t}$ is an $m \times 1$ vector of common factors; $\bm{\gamma_i}$ is the corresponding factor loadings; and $u_{it}$ is the error term. WTI oil price is included as a common factor for all countries. The model is first estimated using LASSO and OCMT separately, followed by post-estimation to obtain the coefficient estimates.

Stacking across all targets yields an $N \times 10N$ subset of the $10N \times 10N$ full coefficient matrix $\hat{\bm{B}}$, with rows indexing targets and columns indexing driver blocks (country-specific and common). The lower $9N$ rows corresponding to variables other than oil demand are set to zero since they are not of interest. The network matrix $\bm{\hat{\kappa}}$ is then constructed as in equation \eqref{eq: concentration matrix}, where each row $i$ in $\bm{\hat{\kappa}}$ is scaled by the inverse of the residual variance, $1/\hat{\sigma}_i^2$. 
To prevent variables with large column norms driven by scaling--but supported only by sparse connections--from being spuriously classified as dominant drivers, we focus on the subset $\bm{\hat{\kappa}_r}$. This subset retains only columns with non-zero counts above the median and diagonal ratios defined by equation \eqref{eq: diagonal ratio} below a threshold set by researchers. Following  \cite{brownlees2021detecting}, we rank the column norms of $\bm{\hat{\kappa}_r}$ in descending order and compute the sequence of consecutive norm ratios. All Variables up to and including the column associated with the largest ratio are identified as dominant drivers. In addition, the marginal share of the $s$-th dominant driver has to be at least $1/\hat{N}_d$ to be retained. Once dominant drivers are selected, an $h$-step forecast is based on:
\begin{equation}
     \Delta \ln OC_{i,t+h} = \alpha_i^h + \rho_{i}^h(L) \Delta \ln OC_{i,t} +  \textbf{dd}_{t}' \bm{\beta}^h_i(L) + u_{i,t+h},
\end{equation}
where $L$ denotes the lag operator ($L^k x_t = x_{t-k}$), $\rho_{i}^h(L)$ and $\bm{\beta}^h_i(L)$ are lag polynomials including contemporaneous term ($L^0$), and $\textbf{dd}_{t}$ denotes the dominant drivers. Dominant drivers selection and model estimation are conducted on the training sample, while predictions are generated on the testing sample.

\subsection{Competitors}
\textbf{AR:}  
A natural benchmark is the autoregressive (AR) model, in which forecasts rely on lagged values of oil consumption growth for each country. The specification is  
\begin{equation}
    \Delta \ln OC_{i,t+h} = \alpha_i^h + \rho_i^h(L)\,\Delta \ln OC_{i,t} + u_{i,t+h},
\end{equation}
This captures purely time-series persistence without relying on external predictors. The \textbf{ARX} model augments this specification by forcing the inclusion of a fixed set of fundamentals: population, GDP, energy intensity, and the global WTI oil price.  

\textbf{$\textbf{LASSO}_i$:}  
We consider a rigorous LASSO approach implemented separately for each country. Unlike the dominant drivers method, which selects a common set of predictors across countries, LASSO performs country-specific variable selection. The $h$-step forecasting equation is  
\begin{equation}
    \Delta \ln OC_{i,t+h} = \alpha_i^h + \rho_i^h(L)\,\Delta \ln OC_{i,t} + \mathbf{X}_{i,t}' \bm{\beta}_i^h(L) + u_{i,t+h},
\end{equation}
where $\mathbf{X}_{i,t}$ is the set of all candidate regressors available for country $i$ including common factors. %This framework allows for heterogeneous predictor sets across countries but does not identify global dominant drivers.

For all methods, lag length is chosen according to the Bayesian Information Criterion (BIC), and the resulting models are used to generate predictions on the testing sample.

%% file: 4Results.tex
\subsection{Dominant drivers selection} \label{}
For the two dominant drivers methods, the first $N$ rows of the network matrix correspond to the countries associated with the dependent variable. All other rows are set to zeros, as we are solely interested in the dominant drivers of oil consumption. However, there are three main ways in which spurious selection of dominant drivers can creep in. First, due to the presence of row-specific residual variance in the denominator of the network matrix, variables with large column norms mainly resulting from small residual variances may be incorrectly identified as dominant drivers. To mitigate this, we count the number of non-zero elements of each column and retain only those with a number of connections exceeding the median. Second, a non-dominant unit with a large diagonal ratio defined in \eqref{eq: diagonal ratio} may be falsely selected, since the diagonal element is simply the inverse of the residual variance. To address this potential issue, we further restrict our attention to columns with diagonal ratios below a reasonable share (50\% - 60\%). Note that this restriction only applies to the first $N$ columns corresponding to the dependent variable. And third, it is possible for a column with a relatively small norm share to exhibit the largest consecutive norm ratio, owing to a sharp decline in the subsequent column’s share to nearly zero. Therefore, a minimum norm share of $1/N$ is imposed.

\begin{figure}[h]
    \centering
    \includegraphics[width=\linewidth]{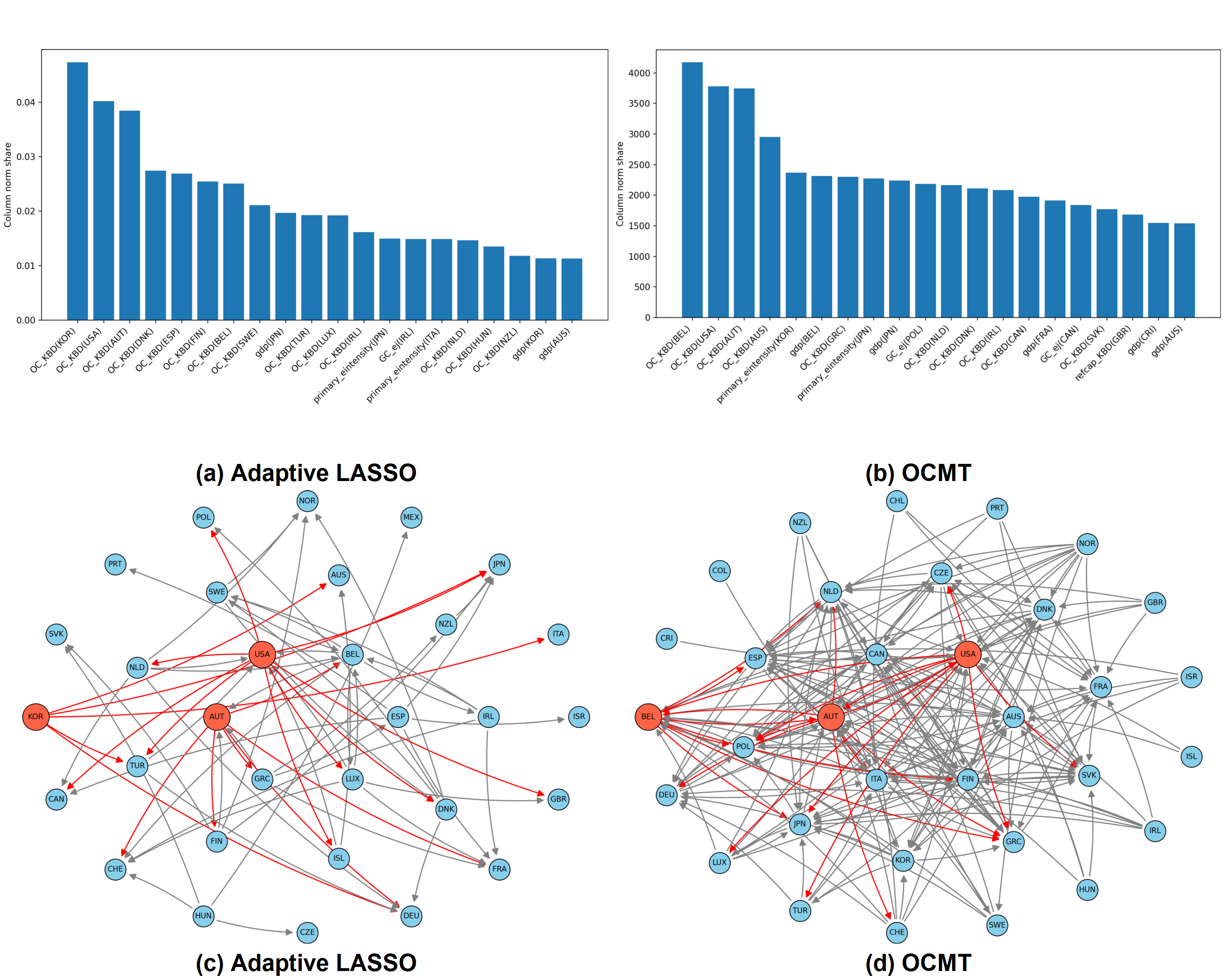}
    \caption{Column norm shares and network graphs - OECD}
    \label{fig: norm and network OECD}
\end{figure}

 Columns of the filtered sub-matrix are subsequently ranked in descending order according to the $L_1$ norm. All units up to and including the column corresponding to the largest consecutive norm ratio are identified as dominant drivers. Figure \ref{fig: norm and network OECD} presents the top 20 sorted column norms of dominant-driver candidates for OECD region, along with the network graphs depicting the relationships between country-level oil demand and the selected dominant drivers. In particular, panels (a) and (b) show that South Korea, the United States, and Austria are identified as dominant drivers by adaptive LASSO, whereas Belgium, the United States, and Austria are selected as dominant drivers by OCMT for the OECD sample. Panels (c) and (d) plot the corresponding directed connections among all OECD countries with red nodes representing identified dominant drivers. Both Austria and the United States are consistently selected as dominant drivers for OECD region, while the disagreement on South Korea and Belgium is expected under high correlation among countries. Normalisation within each country eliminates the effects of scale difference and makes it possible for small economies to be selected as dominant drivers. However, interpreting Austria as a dominant driver is less compelling on purely economic grounds. The links from Austria to Belgium, Switzerland, Germany, Finland, France, Greece, Italy, Netherlands and the United States suggest that Austria can be regarded as a proxy for part of the European common factor.
 
Figure \ref{fig: norm and network OECD PC} also plots the corresponding dominant drivers and network graphs when unobserved common factors are controlled for via CCE. For both approaches, the effect of the United States is absorbed into the common factors, which corroborates the observation that the United States acts as a global dominant driver. Meanwhile, France becomes more representative of a Europe-based factor not fully captured by the global CCE averages and is selected by both methods. We also find that LASSO selects additional dominant drivers—South Korea, Japan, Austria, and Belgium—reflecting comparable but modest gains in predictive performance after controlling for unobserved common factors.
 
\begin{figure}[h]
    \centering
    \includegraphics[width=\linewidth]{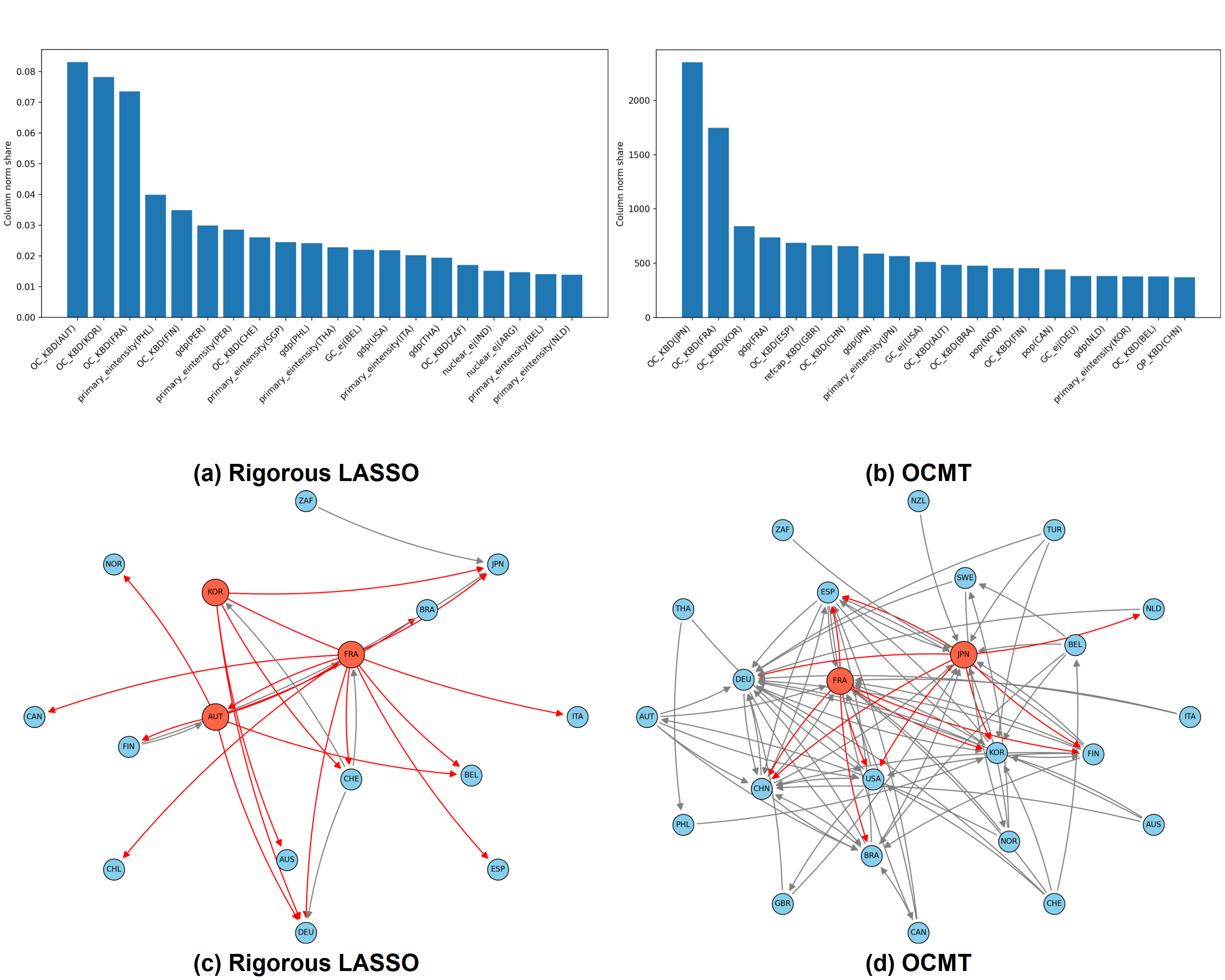}
    \caption{Column norm shares and network graphs - GVAR}
    \label{fig: norm and network GVAR}
\end{figure}

Similarly, Figure \ref{fig: norm and network GVAR} shows the top 20 sorted column norms of dominant-driver candidates and the network graphs for GVAR countries. While the LASSO approach selects Austria, South Korea and France as dominant drivers, the OCMT approach picks up Japan and France. With both approaches, a European hub and an Asian hub are selected. After controlling for unobserved common factors, Figure \ref{fig: norm and network GVAR PC} shows that OCMT still finds France and Japan as key drivers for oil demand, while LASSO adds Belgium as an additional dominant driver. Again, we attribute this to the small but correlated residual patterns among the countries in question which deliver improvements in prediction accuracy even after controlling for common factors.

Taken together, the results identify France and Japan as the robust dominant drivers in the oil demand network for both OECD and GVAR countries. France consistently captures a persistent European factor, while Japan represents an Asian component. Other economies occasionally selected by LASSO (Austria, Belgium, and South Korea) act as redundant proxies that provide only incremental improvements in predictive accuracy.

\subsection{Forecast performance comparisons}
\begin{figure}[htbp]
    \centering
    % First (top) subfigure
    \begin{subfigure}{0.8\textwidth}
        \centering
        \includegraphics[width=\linewidth]{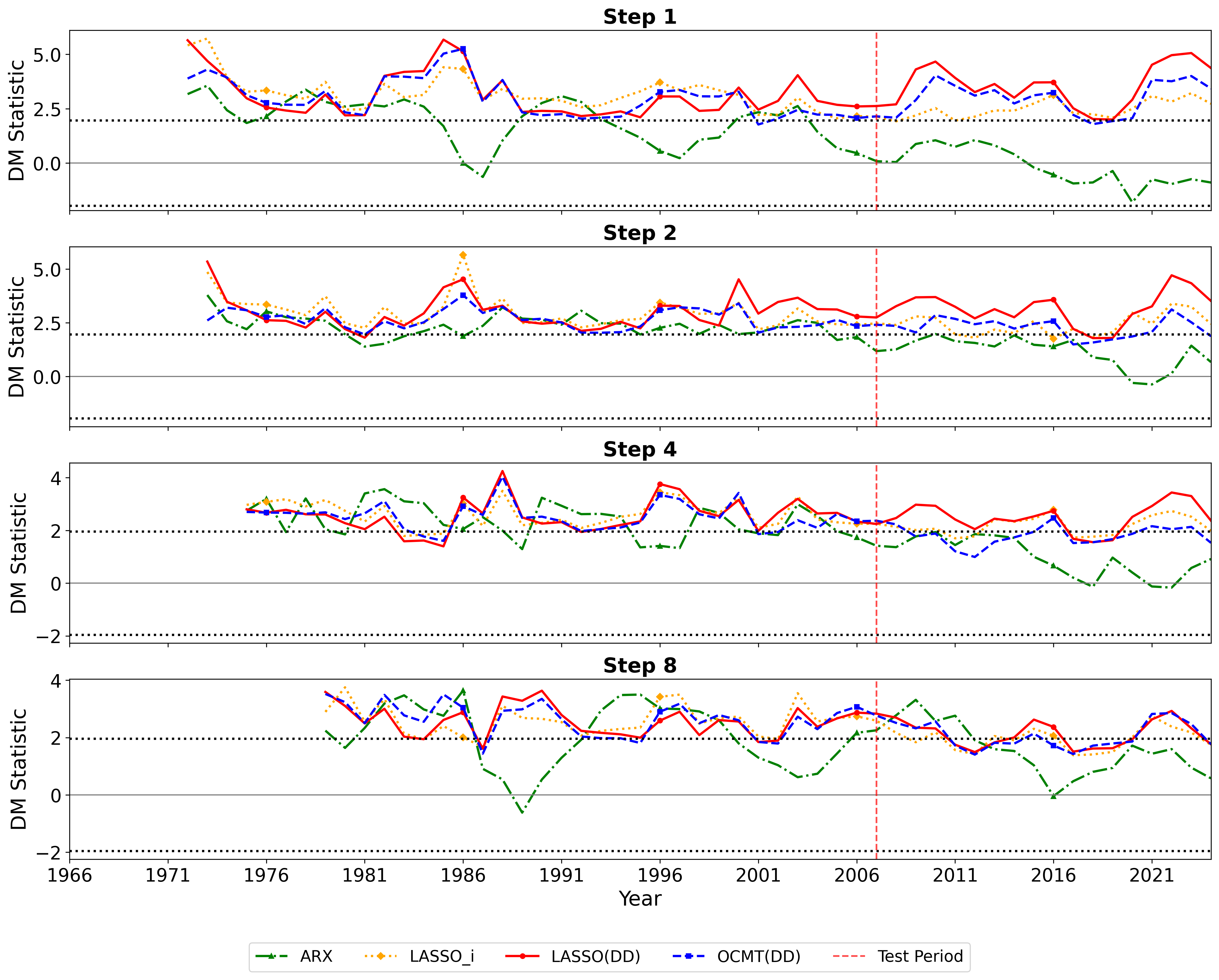}
        \caption{Diebold-Mariano Test statistics - AR benchmark (OECD)}
        \label{fig:DM forecast OECD}
    \end{subfigure}

    \vspace{4mm}

    % Second (bottom) subfigure
    \begin{subfigure}{0.8\textwidth}
        \centering
        \includegraphics[width=\linewidth]{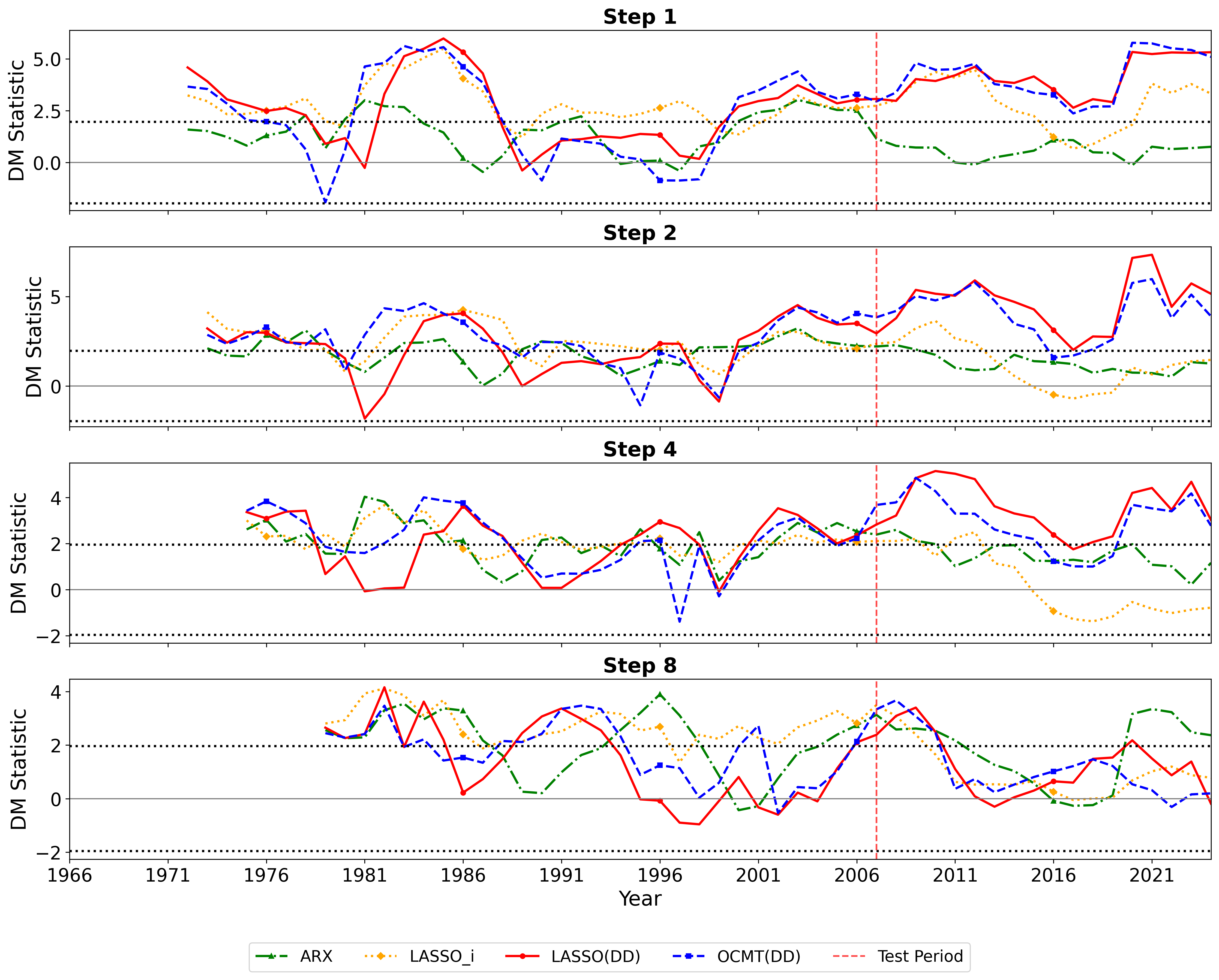}
        \caption{Diebold-Mariano Test statistics - AR benchmark (GVAR)}
        \label{fig:DM forecast GVAR}
    \end{subfigure}

    \caption{Cross-sectional Diebold-Mariano Test statistics – AR benchmark }
    \vspace{2mm}
    \raggedright
    \footnotesize\textit{Note:} Cross-sectional test statistics for comparisons between competitors and the benchmark with the null of equal forecasts. ARX stands for the AR model with additional regressors, LASSO(DD) and OCMT(DD) are dominant-driver methods, $\text{LASSO}_i$ selects each country's own regressors. The horizontal black dashed lines denote the 5\% significance level, and the vertical dashed line separates the training and testing samples at year 2007.
    \par % end raggedright
\end{figure}

Figure \ref{fig:DM forecast OECD} shows the cross-sectional Diebold-Mariano test statistics for several methods, including the two dominant-driver methods: adaptive LASSO and OCMT, the $\text{LASSO}_i$ approach, which selects country-specific regressors for each country, and the AR model augmented with additional fundamental regressors against a standard AR benchmark for OECD countries over forecast horizons $h=1$ to $h=8$. The null hypothesis assumes equal predictive accuracy between the benchmark and the competitor, with positive values indicating that the competitor outperforms the benchmark. The horizontal black dashed lines denote the 5\% significance level, and the vertical dashed line separates the training and testing samples at year 2007. Evidently, all dominant-driver methods and $\text{LASSO}_i$ outperform the benchmark AR model in the testing sample. The dominant drivers set selected by LASSO yields the highest predictive accuracy, followed by OCMT and the $\text{LASSO}_i$ method. Forcing fundamental regressors (population, GDP, energy intensity, and the global WTI oil price) into the AR model does not lead to any improvements in forecasting performance compared to the simple AR. Unsurprisingly, as the forecast horizon increases and uncertainty grows, the performance of all competing methods steadily declines, but evidence suggests that this happens to a lesser extent for the dominant driver-based forecasts, which we propose as part of this study. 

An interesting finding is that dominant driver methods, especially the algorithm based on LASSO, yields remarkably larger and positive DM test statistics during volatile periods such as the 2008 global financial crisis, the pandemic, and geopolitical conflicts in recent years, all of which caused large shocks to global oil demand. Intuitively, dominant drivers absorb and transmit shocks better than non-dominant drivers. Including dominant drivers in the model provides cleaner signals for the global oil demand prediction than including each country's own specific but noisy information, especially in the presence of unexpected global shocks. The pattern also supports the view that the selected dominant units are indeed structurally important to global oil demand dynamics. As shown in Figure \ref{fig:DM forecast GVAR}, the pattern is even more prominent for GVAR countries which cover over more than 90\% of world GDP.\footnote{See Appendix Tables \ref{tab:forecast_oecd_NO_PC}, \ref{tab:forecast_gvar_NO_PC}, \ref{tab:forecast_oecd_PC}, and \ref{tab:forecast_gvar_PC} for the evaluation metrics for all models, including RMSE, MAE, and the Model Confidence Set (MCS) share.}   

Table \ref{table:DM_OECD no factors} reports shares of countries with positive, negative, and significant DM test statistics across all methods. For the four forecast horizons, the LASSO-based dominant drivers method yields positive (significant) DM test statistics in 100\% (96.67\%), 96.67\% (93.33\%), 86.67\% (76.67\%), and 80\% (63.33\%) of OECD countries, where figures in parentheses show the share significant at the 5\% level. The OCMT-based dominant drivers method has marginally fewer but comparable positive shares to the LASSO approach for horizons 1, 2, and 4, and slightly higher for horizon 8. The country-specific $\text{LASSO}_i$ approach has 6.67\%, 10\%, and 3.33\% lower positive shares than the LASSO(DD) approach for horizons 1, 2, and 8. The pattern for GVAR region shown in Table \ref{table:DM_GVAR no factors} is similar, with LASSO(DD) dominating in the time series DM test, followed closely by OCMT(DD) then $\text{LASSO}_i$.

\input{Tables/DM_Table_OECD_NO_F}

\clearpage

\subsection{Robustness check}
While imposing restrictions on the identification of dominant drivers can substantially reduce the likelihood of selecting non-dominant units, it may also undermine out of sample predictive power by excluding potentially relevant candidates. To see whether such restrictions have considerable impacts on the forecast performance of the dominant drivers method, we also compare dominant drivers with and without restrictions in both selection and prediction stages.

As in Figures \ref{fig: norm and network OECD (U)} and \ref{fig: norm and network GVAR (U)}, when the restriction on the diagonal ratio is lifted, the best performing dominant driver for OECD countries selected by LASSO and OCMT are GDP of Japan and oil consumption of Japan, respectively. For GVAR countries, dominant drivers identified by LASSO become oil consumption of Japan, South Korea, and energy intensity of Singapore, which is selected solely due to its large impact on its own oil demand. And the dominant driver selected by OCMT is South Korea. This implies that aggregate oil demand is solely driven by Asian hubs. At face value, this is difficult to reconcile purely with economic intuition, but we see logistical reasons Japan, South Korea, and Singapore could act as drivers. One such reason is that oil consumption in these important countries are imported. As such, their oil demand patterns have implications for the wider oil consumption and production system as well as the way in which substitutes of oil (coal, gas, nuclear, and renewables among others) are managed. Furthermore, Singapore acts as a central hub for oil refinery activity. Crude oil is imported into the country, refined, and re-exported to a wide range of destinations for final consumption. These strong linkages leave traces in the dataset which appear to have been picked up in this context. 

In terms of prediction, Figure \ref{fig: DM forecast Both Unrestricted} shows that relaxing the restriction on the diagonal ratio does not improve forecast performance over the proposed restriction for both OECD and GVAR regions. Time series DM test statistics reported in Tables \ref{table: DM OECD Factors} and \ref{table: DM GVAR Factors} for both regions after controlling for unobserved factors also show no gains and sometimes losses in terms of shares of countries outperforming the AR model. Specifically, the positive shares of unrestricted OCMT approach are lower than the restricted OCMT by 6.67\% and 10\% for horizons 2 and 8 in OECD region, respectively. In the LASSO case, the positive shares of unrestricted LASSO are lower than the restricted one by 3.03\% and 21.21\% for horizons 2 and 8 in GVAR region, respectively. And similar losses are observed for the OCMT approach as well.

\begin{figure}[htbp]
    \centering
    \begin{subfigure}{0.48\textwidth}
        \centering
        \includegraphics[width=\linewidth]{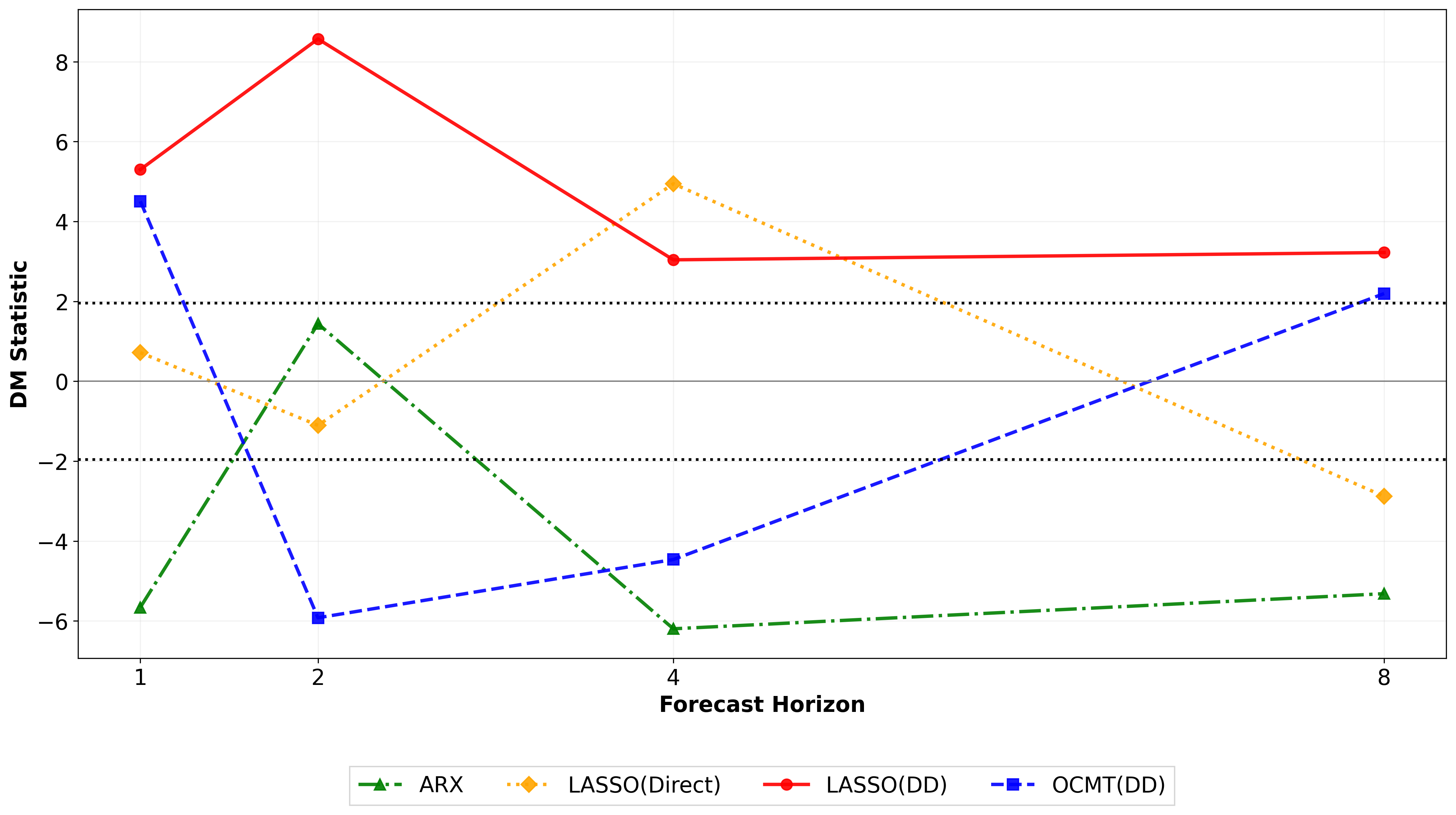}
        \caption{OECD}
        \label{fig: total OECD}
    \end{subfigure}\hfill
    \begin{subfigure}{0.48\textwidth}
        \centering
        \includegraphics[width=\linewidth]{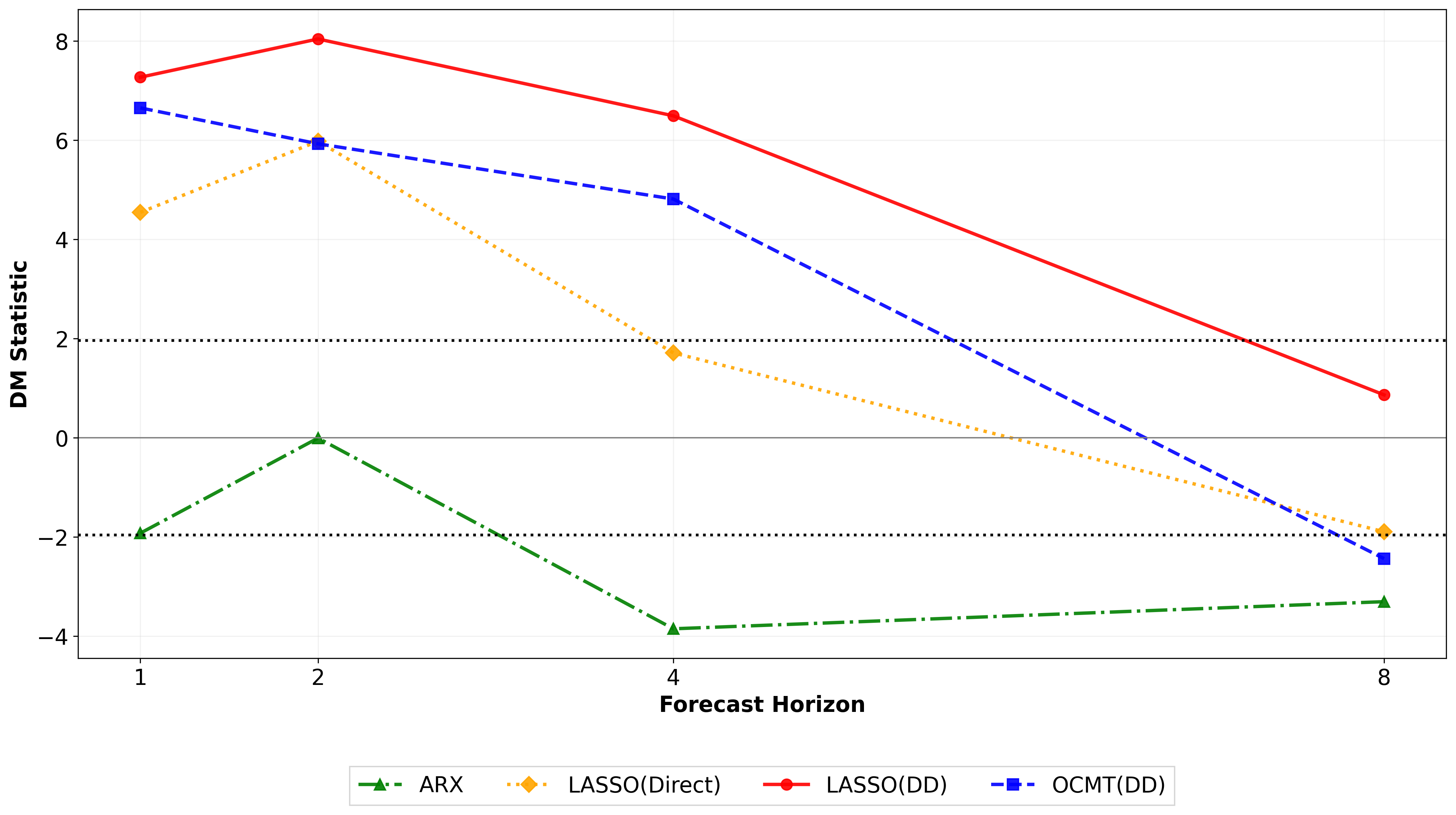}
        \caption{GVAR}
        \label{fig: total GVAR}
    \end{subfigure}
    \caption{Time series Diebold-Mariano test statistics for aggregate oil consumption - AR benchmark}
    \label{fig:total}
        \vspace{2mm}
    \raggedright
    \footnotesize\textit{Note:} Time series test statistics for comparisons between competitors and the benchmark with the null of equal forecasts. ARX stands for the AR model with additional regressors, LASSO(DD) and OCMT(DD) are dominant-driver methods, RLASSO stands for rigorous LASSO that directly selects regressors for aggregate oil demand. The horizontal dashed lines denote the 5\% significance level.
    \par % end raggedright
\end{figure}

In addition, the selected dominant drivers are also used to predict the growth rate of aggregate oil consumption defined in \eqref{eq: agg oil cons} across horizons and regions as shown in Figure \ref{fig:total}. The two dominant-driver methods are comparable in most specifications. The LASSO(DD) approach significantly outperforms the benchmark across all horizons for OECD countries and at horizons 1–4 for GVAR countries, while the OCMT(DD) approach outperforms the benchmark at horizons 1 and 8 for OECD countries and at horizons 1–4 for GVAR countries. A rigorous LASSO that directly selects regressors for aggregate oil demand is included as a competitor, which is almost always outperformed by our dominant-driver approach. Forcing important regressors into the AR model reduces the predictive performance. 

In summary, the dominant drivers selected by our proposed method are economically reasonable and  deliver robust improvement in predicting both country-level and regional aggregate oil demand for OECD and GVAR regions.

\subsection{SVAR results}
We study the dynamic transmission of dominant-driver oil demand shocks to aggregate oil demand growth using a structural VAR (SVAR) framework. For each specification, the VAR includes an aggregate oil demand growth series and the set of corresponding dominant drivers identified by LASSO and OCMT for OECD and GVAR countries respectively. Let $\bm{y_t}$ denote the $S \times 1$ vector of endogenous variables, where $S$ equals one plus the number of dominant drivers, in a given specification. The reduced form VAR is

\begin{equation} \label{eq: reduced-VAR}
    \bm{y_t} = \bm{c} + \sum_{l=1}^p \bm{A_l y_{t-l}} + \bm{u_t}, \quad \bm{u_t} \sim N(0,\bm{\Sigma_u}),      
\end{equation}
where $\bm{c}$ is a vector of constants and $p \leq 2$ is the lag length selected by the Akaike information criterion (AIC). The bound is imposed to preserve degrees of freedom given the annual frequency and the multivariate nature of the system.

The Cholesky identification scheme is used to identify structural shocks. In particular, the reduced form errors are assumed to admit a contemporaneous decomposition:

\begin{equation} \label{eq: SVAR error decom}
    \bm{u_t = B \epsilon_t}, \quad \mathbb{E}(\bm{\epsilon_t \epsilon_t'}) = \bm{I}_S,
\end{equation}
where $\bm{B}$ is an $S \times S$ lower triangular matrix obtained via the Cholesky decomposition of $\bm{\Sigma_u}$ and $\epsilon_t$ are structural orthogonal shocks. Thus, the economic identification of structural shocks relies entirely on the variable ordering. In all specifications, the aggregate oil demand growth is ordered first. Therefore, we implicitly assume that orthogonalised shocks originating from dominant drivers have no contemporaneous effects on the aggregate demand. The baseline ordering for each specification is reported in Table \ref{tab:svar_specs}. All estimated reduced form VARs satisfy the standard stability condition since the maximum modulus of the companion matrix roots is below one in every specification.
\input{Tables/SVAR}

\begin{figure}[htbp]
    \centering
    \includegraphics[width=0.9\linewidth]{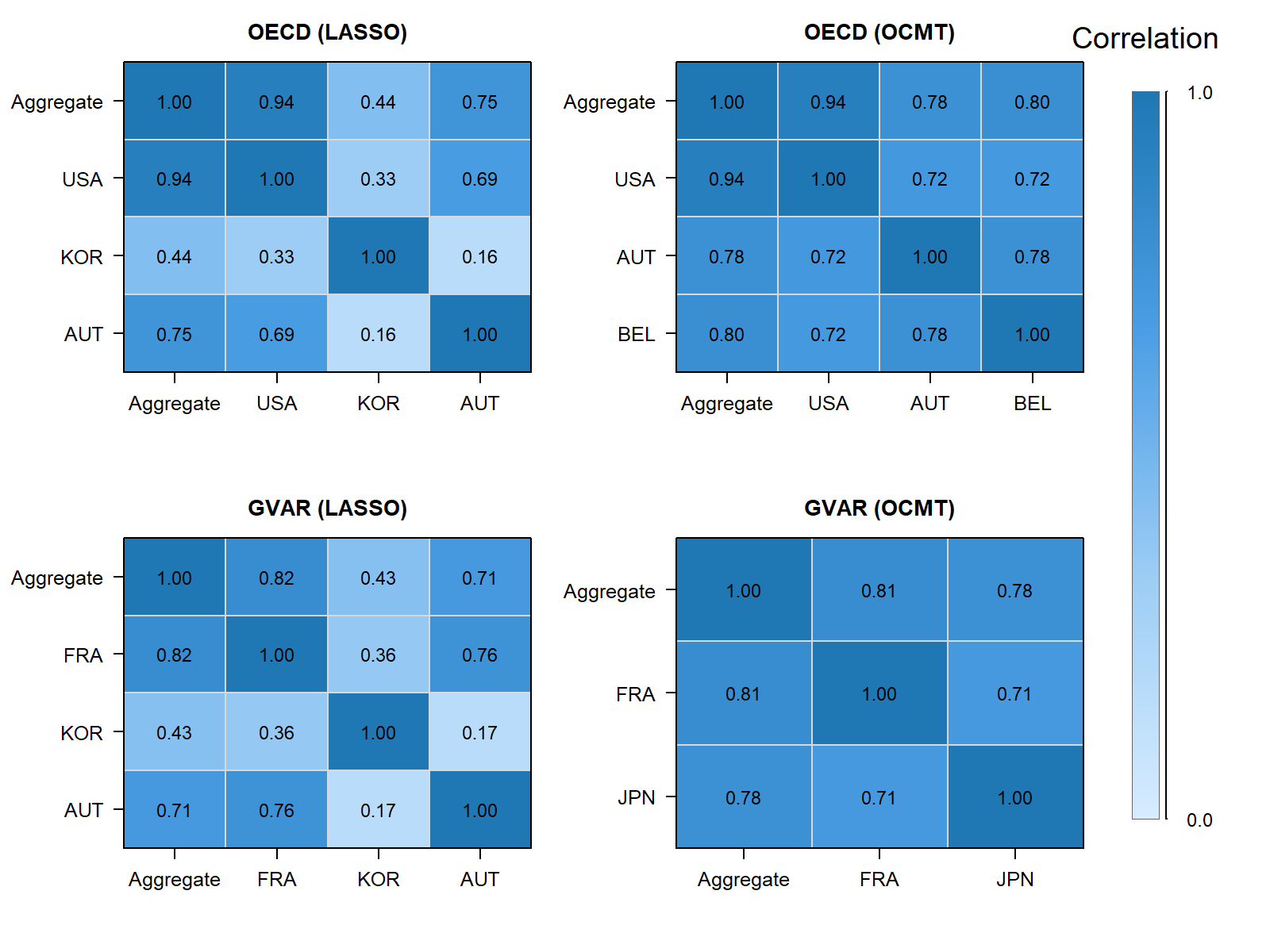}
    \caption{Contemporaneous correlation matrices in all specifications}
    \label{fig:error-corr-svar}
\end{figure}

Because results can be sensitive to dominant-driver orderings when reduced form errors are highly contemporaneously correlated as shown in Figure \ref{fig:error-corr-svar}. Therefore, all possible orderings are evaluated as part of robustness analysis. Due to space constraints, results for the baseline Cholesky ordering in Table \ref{tab:svar_specs} are reported. Alternative orderings yield qualitatively similar impulse-response and variance decomposition patterns. We summarise this robustness using the range of FEVD shares at the 10-year horizon (“Range@10”) reported below in Table \ref{tab:fevd_aggregate}.

Impulse response functions (IRFs) are computed for a horizon of 10 years in each specification. 95\% Confidence intervals (CI) are obtained using bootstrap resampling of the estimated VAR with 500 replications. Since aggregate oil demand is ordered first, innovations to dominant drivers have zero contemporaneous effects on aggregate oil demand at horizon $h=0$. 

\begin{figure}[htbp]
    \centering

    \begin{subfigure}{0.8\linewidth}
        \centering
        \includegraphics[width=\linewidth]{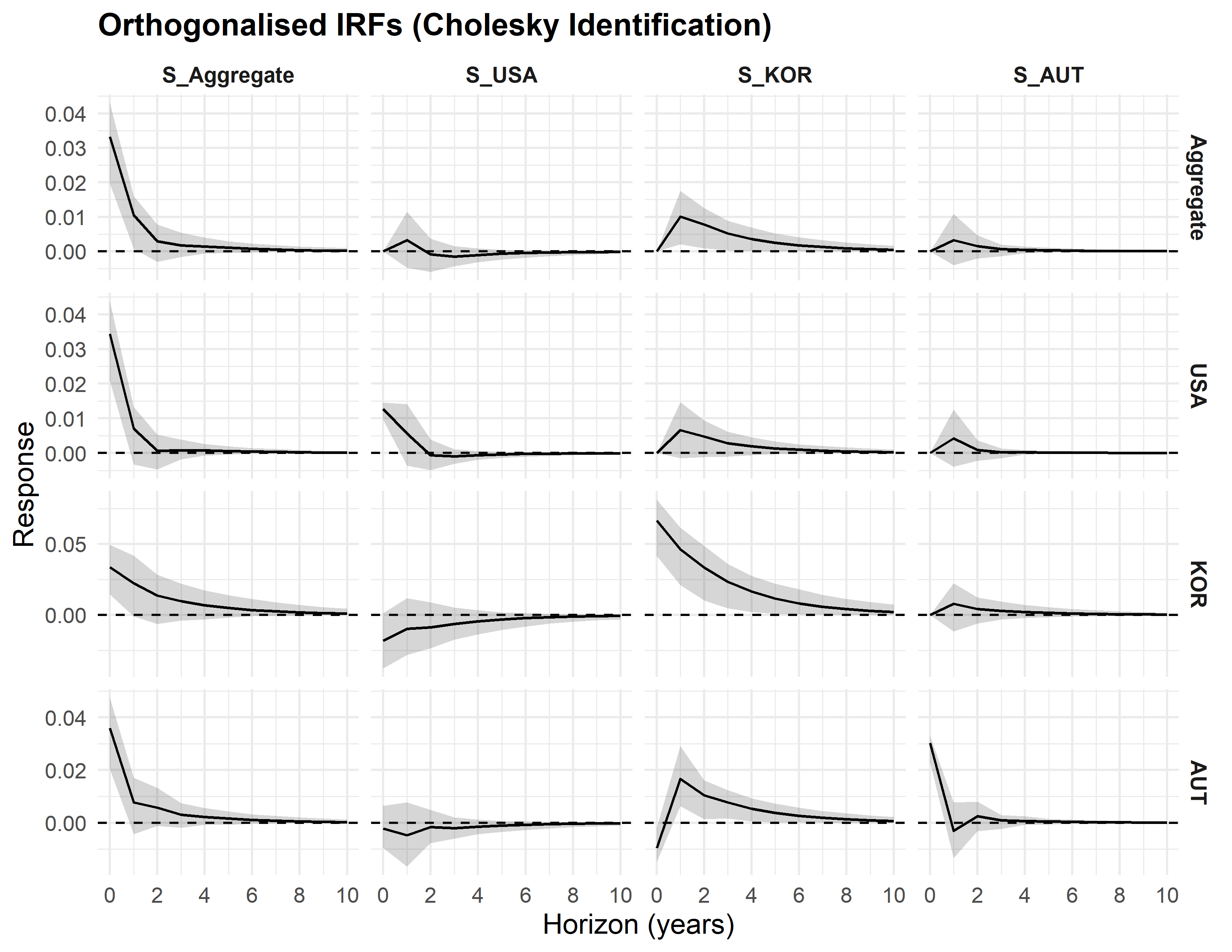}
        \caption{LASSO}
        \label{fig:irf-lasso-oecd}
    \end{subfigure}

    \vspace{0.6em}

    \begin{subfigure}{0.8\linewidth}
        \centering
        \includegraphics[width=\linewidth]{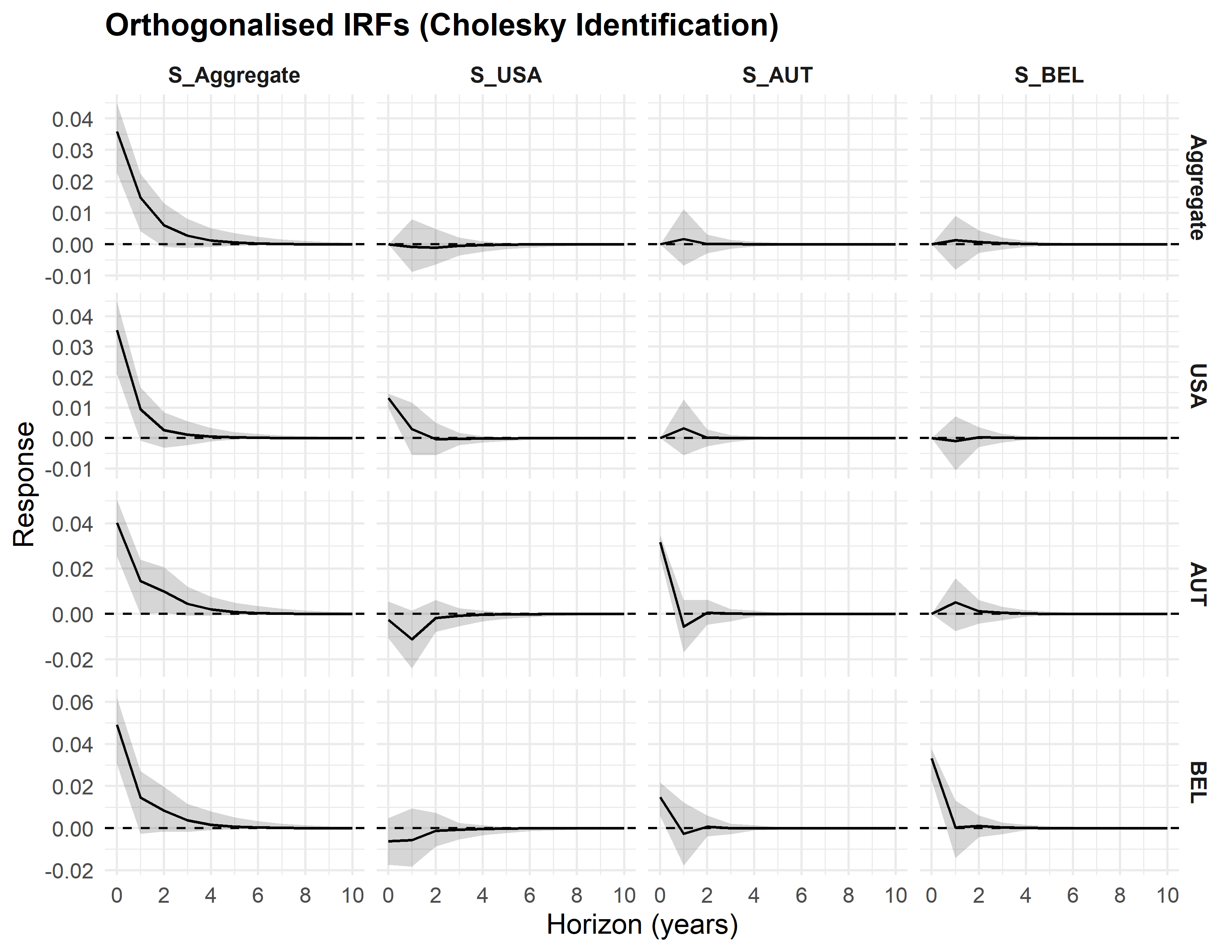}
        \caption{OCMT}
        \label{fig:irf-ocmt-oecd}
    \end{subfigure}

    \caption{Recursive identification (Cholesky) for the OECD sample. Columns (S\_$\cdot$) are orthogonalised shocks and rows are responses. The aggregate series is ordered first, followed by dominant-driver countries in the order shown in the panels. 95\% confidence bands with 500 bootstrap replications. VAR lag length (1) selected by AIC.}
    \label{fig:irf-oecd}
\end{figure}

\begin{figure}[htbp]
    \centering

    \begin{subfigure}{0.8\linewidth}
        \centering
        \includegraphics[width=\linewidth]{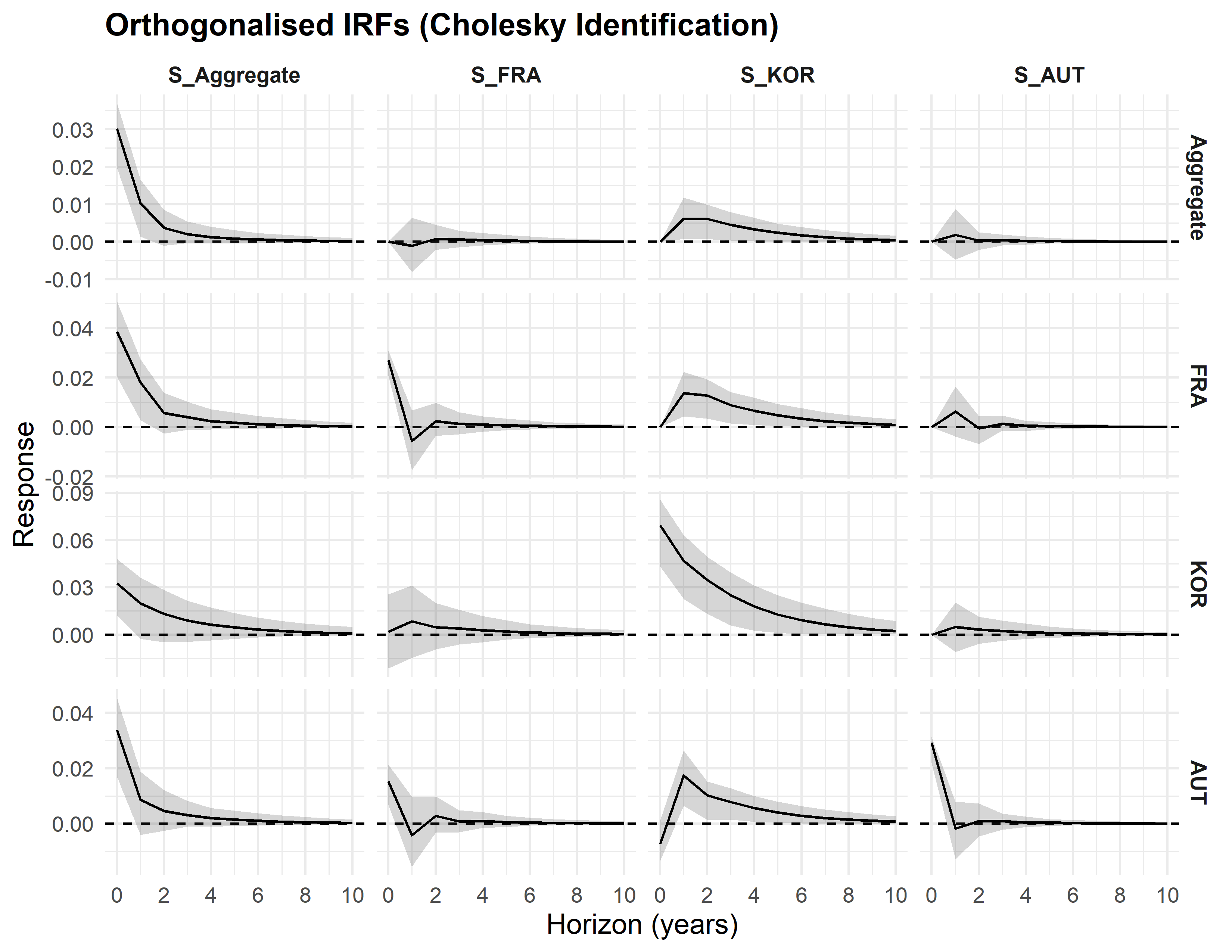}
        \caption{LASSO}
        \label{fig:irf-lasso-gvar}
    \end{subfigure}

    \vspace{0.6em}

    \begin{subfigure}{0.8\linewidth}
        \centering
        \includegraphics[width=\linewidth]{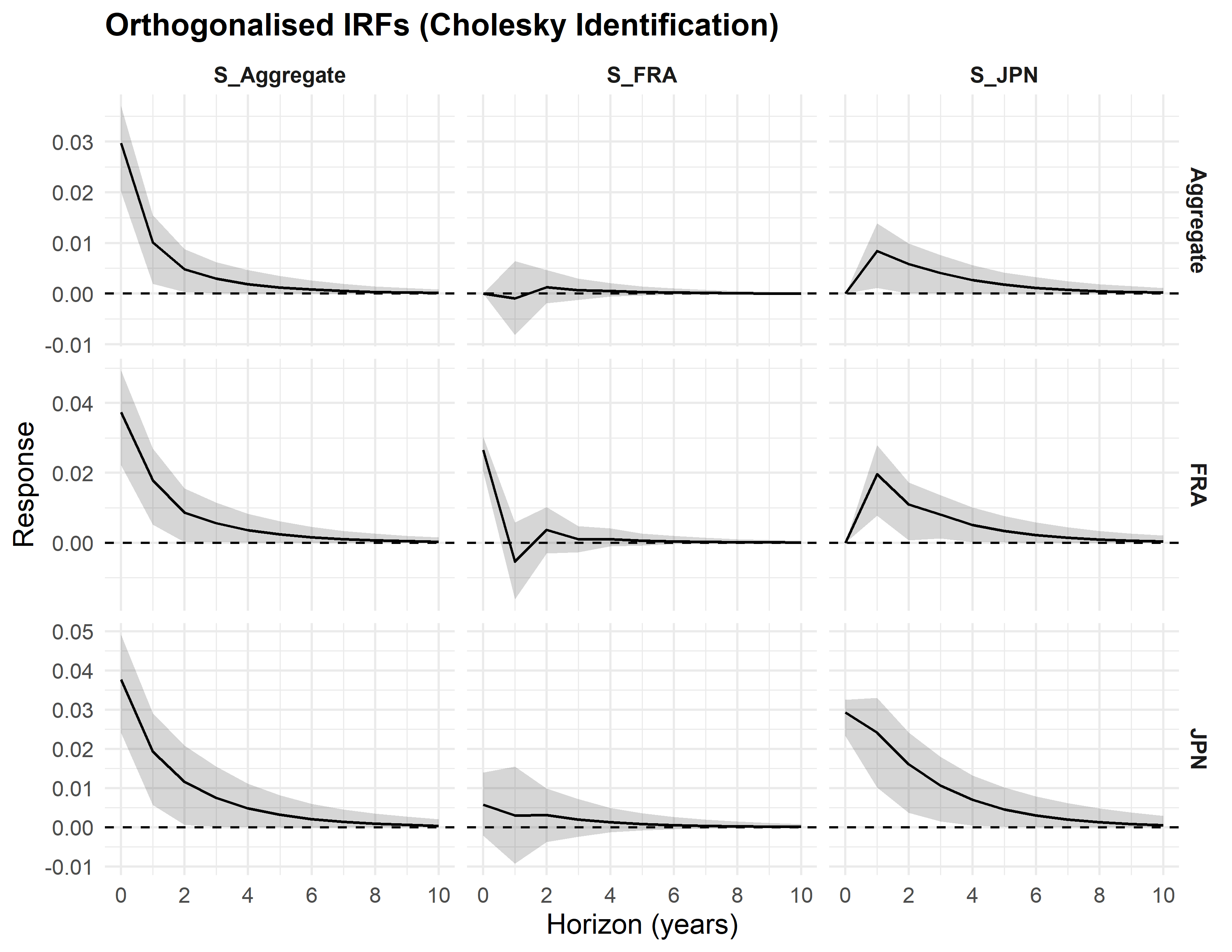}
        \caption{OCMT}
        \label{fig:irf-ocmt-gvar}
    \end{subfigure}

    \caption{Recursive identification (Cholesky) for the GVAR sample. Columns (S\_$\cdot$) are orthogonalised shocks and rows are responses. The aggregate series is ordered first, followed by dominant-driver countries in the order shown in the panels. 95\% confidence bands with 500 bootstrap replications. VAR lag length (1) selected by AIC.}
    \label{fig:irf-gvar}
\end{figure}

The first column in Figures \ref{fig:irf-oecd}-\ref{fig:irf-gvar} finds that a global demand shock affect positively aggregate and all countries oil consumptions. Moreover, Figure \ref{fig:irf-lasso-oecd} shows that, with dominant drivers selected by LASSO, a shock to Korea produces a positive transmission to OECD Aggregate oil demand growth. The response peaks at $h=1$ with value 0.0101 and a 95\% bootstrap CI [0.0020, 0.0176], and remains statistically significant thereafter. The effects then decay monotonically and become close to zero by the medium run. Overall, the response indicates a short-run amplification followed by gradual mean reversion. Intuitively, an oil-demand innovation in a key manufacturing and trading hub is informative about broader conditions an supply-chain activity that may propagate to aggregate demand, but the effect gradually fades away as activity stabilises. While shocks to the USA and Austria show positive but insignificant impacts at 95\% level in general, suggesting imprecise and weaker transmission than Korea. 
Forecast error variance decomposition (FEVD) in Table \ref{tab:fevd_aggregate} confirms that Korea accounts for 14.57\% of aggregate demand forecast error variance by $h=10$, followed by USA (1.08\%) and Austria (0.91\%).
On the contrary, Table \ref{fig:irf-ocmt-oecd} shows the dominant drivers selected by OCMT contribute only marginally to OECD aggregate oil demand fluctuations. This is consistent with the forecasting evidence in Figure \ref{fig: total OECD} where LASSO approach yields clear gains over an AR benchmark while the OCMT method does not.

Similarly, Figure \ref{fig:irf-lasso-gvar} presents that GVAR aggregate oil demand responds positively and most strongly to Korea. The response peaks at $h=1$ with value 0.0061 and 95\% CI [0.0008, 0.0118] and gradually approaches zero thereafter. The response to an Austria shock is positive, whereas the response to a France shock is slightly negative initially. From Figure \ref{fig:irf-ocmt-gvar}, we observe that the aggregate responds positively to an orthogonalised shock to Japan, which peaks at $h=1$ with value 0.0084 and bootstrapped CI [0.0014, 0.0139] and remains significant through $h=3$. The aggregate response to a France shock is again slightly negative initially but reverts to positive quickly. The FEVD in Table \ref{tab:fevd_aggregate} at $h=10$ shows that Korea explains about 10.08\% of GVAR aggregate oil demand forecast error variance, compared with 0.21\% by France and 0.35\% by Austria. With Japan and France identified as dominant drivers, FEVD at $h=10$ shows that Japan accounts for 11.58\% of the forecast error variance, followed by France with 0.29\%.

\input{Tables/FEVD}

Putting together, the results point to an Asian hub (Korea or Japan) as a dominant transmitter of oil-demand shocks to both OECD and GVAR aggregate oil demand growth over 1966–2024. Economically, this is plausible because these economies are highly integrated into global manufacturing and trade networks and have comparatively oil-intensive industrial and transport sectors. Consequently, country-level demand innovations are more likely to proxy for shifts in global activity. While other countries’ shocks may be either more idiosyncratic or more strongly collinear with the hub innovation, once orthogonalised they contribute only marginal variance to the aggregate.

To ensure that this conclusion is not driven by a specific Cholesky ordering among dominant drivers, we perform the analysis for all permutations and record the range of FEVD shares for each dominant driver at $h=10$ (``Range$@10$''). Korea accounts for 10.08\% - 15.29\% and Japan explains 11.32\% - 11.58\% of the forecast error variance decomposition across all orderings.

Finally, because dominant drivers are included mechanically in the construction of the aggregate series, we also construct an alternative aggregate oil demand growth series that exclude the dominant drivers in each specification. Results in Figures \ref{fig:irf-oecd-exc} and \ref{fig:irf-GVAR-exc} exhibit similar patterns to the baseline results, suggesting that the identified transmission is not driven by mechanical aggregation but by real dynamic propagation from dominant-driver innovations.

%% file: Tables/DM_Table_OECD_NO_F.tex
\begin{table}[htbp]
\tiny
\centering
\resizebox{\textwidth}{!}{
\begin{tabular}{@{\extracolsep{4pt}}l cccc @{}}
\hline
\hline
OECD & ARX & LASSO\(_i\) & LASSO(DD) & OCMT(DD) \\
\hline
\multicolumn{5}{l}{1 Step} \\
\hline
\(1/34 \sum_{i=1}^N I(S_{T,i} \leq 0)\) & 30.00 & 6.67 & 0.00 & 3.33 \\
sig. & 20.00 & 0.00 & 0.00 & 0.00 \\
\(1/34 \sum_{i=1}^N I(S_{T,i} > 0)\) & 70.00 & 93.33 & 100.00 & 96.67 \\
sig. & 63.33 & 86.67 & 96.67 & 83.33 \\
\hline
\multicolumn{5}{l}{2 Step} \\
\hline
\(1/34 \sum_{i=1}^N I(S_{T,i} \leq 0)\) & 40.00 & 13.33 & 3.33 & 10.00 \\
sig. & 23.33 & 3.33 & 0.00 & 3.33 \\
\(1/34 \sum_{i=1}^N I(S_{T,i} > 0)\) & 60.00 & 86.67 & 96.67 & 90.00 \\
sig. & 60.00 & 86.67 & 93.33 & 70.00 \\
\hline
\multicolumn{5}{l}{4 Step} \\
\hline
\(1/34 \sum_{i=1}^N I(S_{T,i} \leq 0)\) & 40.00 & 13.33 & 13.33 & 16.67 \\
sig. & 16.67 & 6.67 & 6.67 & 10.00 \\
\(1/34 \sum_{i=1}^N I(S_{T,i} > 0)\) & 60.00 & 86.67 & 86.67 & 83.33 \\
sig. & 46.67 & 73.33 & 76.67 & 46.67 \\
\hline
\multicolumn{5}{l}{8 Step} \\
\hline
\(1/34 \sum_{i=1}^N I(S_{T,i} \leq 0)\) & 30.00 & 23.33 & 20.00 & 16.67 \\
sig. & 6.67 & 10.00 & 6.67 & 3.33 \\
\(1/34 \sum_{i=1}^N I(S_{T,i} > 0)\) & 70.00 & 76.67 & 80.00 & 83.33 \\
sig. & 56.67 & 66.67 & 63.33 & 70.00 \\
\hline
\hline
\end{tabular}}
\caption{Test of equal forecast accuracy against a simple AR. \(S_{T,i}\) is the share of countries with a positive or negative test statistic from the Diebold-Mariano test \citep{Diebold1995}. ``Sig.'' indicates the percentage share of cases where \(S_{T,i}\) are significantly positive or negative at the 5\% level. A positive \(S_{T,i}\) implies that forecasts based on dominant drivers are more accurate. ARX is the AR model with forced fundamentals included as regressors, $\text{LASSO}_i$ is the country specific variable selection method, LASSO(DD) and OCMT(DD) are dominant-driver methods with restricted diagonal ratio.}
\label{table:DM_OECD no factors}
\end{table}

%% file: Tables/SVAR.tex
\begin{table}[!htbp]
\centering
\caption{SVAR specifications for regional oil demand growth (Cholesky ordering)}
\label{tab:svar_specs}

\begin{tabular}{l l c c}
\hline
Specification & Cholesky ordering & $p$ (AIC) & Max $|\text{root}|$ \\
\hline
OECD (LASSO) & Aggregate $\rightarrow$ USA $\rightarrow$ KOR $\rightarrow$ AUT & 1 & 0.7054 \\
OECD (OCMT)  & Aggregate $\rightarrow$ USA $\rightarrow$ AUT $\rightarrow$ BEL & 1 & 0.4547 \\
GVAR (LASSO) & Aggregate $\rightarrow$ FRA $\rightarrow$ KOR $\rightarrow$ AUT & 1 & 0.7147 \\
GVAR (OCMT)  & Aggregate $\rightarrow$ FRA $\rightarrow$ JPN & 1 & 0.6547 \\
\hline
\end{tabular}

\vspace{0.3em}
\begin{minipage}{0.97\linewidth}
\footnotesize
\textit{Notes:} $p$ is the number of lags selected by AIC (max lag = 2). Max $|\text{root}|$ is the maximum modulus of the roots of the VAR characteristic polynomial. Values below one indicate a stable VAR.
\end{minipage}
\end{table}

%% file: Tables/FEVD.tex
\begin{table}[!htbp]
\centering
\caption{FEVD of Aggregate forecast error variance (\%)}
\label{tab:fevd_aggregate}
\small
\begin{tabular}{l r r r r r}
\hline
Specification & Horizon & Aggregate & Driver 1 & Driver 2 & Driver 3 \\
\hline
OECD (LASSO) & 1  & 100.00 & 0.00 (USA) & 0.00 (KOR) & 0.00 (AUT) \\
            & 2  & 90.77 & 0.81 (USA) & 7.64 (KOR) & 0.78 (AUT) \\
            & 5  & 84.18 & 1.02 (USA) & 13.89 (KOR) & 0.91 (AUT) \\
            & 10 & 83.45 & 1.08 (USA) & 14.57 (KOR) & 0.91 (AUT) \\
            & Range@10 & --- & \texttt{[1.08--2.79]} & \texttt{[13.24--15.29]} & \texttt{[0.16--0.91]} \\
\hline
OECD (OCMT) & 1  & 100.00 & 0.00 (USA) & 0.00 (AUT) & 0.00 (BEL) \\
           & 2  & 99.64 & 0.05 (USA) & 0.20 (AUT) & 0.11 (BEL) \\
           & 5  & 99.52 & 0.14 (USA) & 0.19 (AUT) & 0.15 (BEL) \\
           & 10 & 99.52 & 0.14 (USA) & 0.19 (AUT) & 0.15 (BEL) \\
           & Range@10 & --- & \texttt{[0.08--0.14]} & \texttt{[0.07--0.21]} & \texttt{[0.15--0.32]} \\
\hline
GVAR (LASSO) & 1  & 100.00 & 0.00 (FRA) & 0.00 (KOR) & 0.00 (AUT) \\
            & 2  & 96.07 & 0.10 (FRA) & 3.49 (KOR) & 0.34 (AUT) \\
            & 5  & 90.26 & 0.19 (FRA) & 9.20 (KOR) & 0.35 (AUT) \\
            & 10 & 89.36 & 0.21 (FRA) & 10.08 (KOR) & 0.35 (AUT) \\
            & Range@10 & --- & \texttt{[0.20--0.35]} & \texttt{[10.08--10.24]} & \texttt{[0.06--0.35]} \\
\hline
GVAR (OCMT) & 1  & 100.00 & 0.00 (FRA) & 0.00 (JPN) & --- \\
           & 2  & 93.24 & 0.08 (FRA) & 6.68 (JPN) & --- \\
           & 5  & 88.53 & 0.28 (FRA) & 11.19 (JPN) & --- \\
           & 10 & 88.13 & 0.29 (FRA) & 11.58 (JPN) & --- \\
           & Range@10 & --- & \texttt{[0.29--0.56]} & \texttt{[11.32--11.58]} & --- \\
\hline
\end{tabular}

\vspace{0.3em}
\begin{minipage}{0.97\linewidth}
\footnotesize
\textit{Notes:} ``Range@10'' is the range of FEVD shares at horizon 10 across alternative dominant-driver orderings within the same specification.
\end{minipage}
\end{table}

%% file: 5Summary.tex
This paper proposes to forecast country-level oil demand with dominant drivers detected via an estimated non-symmetric concentration matrix, which is estimated row by row with two high-dimensional selection methods, LASSO and OCMT. The number of dominant drivers is decided by the position of the column with the largest consecutive column norm ratio in the matrix sorted by column norm in descending order. Therefore, we focus on a subset matrix where columns satisfying the diagonal ratio, the number of connections, and the minimum norm share restrictions in Section \ref{sec: DD method} to alleviate potential false selection issue where pseudo dominant drivers with weak connections get selected only because of a small residual variance. The selected dominant drivers are then included as predictors of the oil demand of all countries. 

The networks estimated by both selection approaches show a clear and stable network structure. For both OECD and GVAR samples, the United States, a European hub, and an Asian hub emerge as robust dominant drivers. The United States acts as a global driver whose effect is largely absorbed by common factors once CCE-type cross-sectional averages are included, which is consistent with its central role in the world economy. France and Japan capture persistent regional effects even after controlling the global averages, with France representing an European factor and Japan representing an Asian factor. Other countries such as Austria, Belgium, and Korea are selected only in some specifications by LASSO and appear to serve mainly as additional proxies for these regional components that provide small gains in prediction accuracy rather than as truly independent dominant drivers. Analysing the dynamic transmission on aggregate oil consumption, we find evidence for an Asian hub consisting of Japan or Korea. 

The forecast comparison confirms the value of including global dominant drivers as predictors for all countries, especially during volatile periods. Across horizons and for both OECD and GVAR regions, all dominant-driver models provide significant improvements over a standard AR benchmark. The dominant drivers found by LASSO deliver the largest gains in predictive accuracy, closely followed by OCMT, while the country-specific $\text{LASSO}_i$ performs slightly worse but still better than the AR model. Forcing standard macro fundamentals to the AR model does not improve, suggesting that the most useful information for predicting oil demand is contained in a small set of dominant units rather than in local regressors. The gains are especially strong during volatile periods of large global shocks, such as the 2008 financial crisis, the pandemic, and recent geopolitical conflicts, where models with dominant drivers show clear improvement over both the AR model and the country-specific LASSO approach. Intuitively, including global dominant drivers in the model for all countries provides cleaner signals for the global oil demand prediction than including each country's own selected regressors, especially in the presence of unexpected shocks when countries are highly correlated. 

The robustness checks unveil that the restrictions placed on the diagonal ratio of the BM procedure ensure an economically meaningful interpretation. Once the restrictions are lifted, the procedure tends to select either Asian hubs alone, or variables that are mechanically tied to the unit under consideration (e.g., Japan’s GDP, or Singapore’s energy intensity), which is inconsistent with basic facts of the world economy. In addition, while the unrestricted procedure delivers marginal forecast improvements, the Diebold–Mariano tests generally favour the restricted procedure.

Overall, the above results suggest that a small set of dominant countries can represent the main drivers of global oil demand and provide strong signals for forecasting than models that fit each country individually. The framework is flexible and can be applied to any variable with spatial dependence or network structure. Future research could allow for time-varying networks and dominant drivers as the pattern may change over time. It would also be interesting to investigate how these patterns differ across subregions defined by geography and income.

%% file: 6Appendix.tex
\newpage

\section*{Appendix A: Lists of countries}
\renewcommand{\theequation}{A.\arabic{equation}}
\renewcommand{\thesection}{A}
\setcounter{equation}{0}
\input{Tables/OECD}
\input{Tables/GVAR}

\section*{Appendix B: Dominant drivers selection flow}
\renewcommand{\theequation}{B.\arabic{equation}}
\renewcommand{\thesection}{B}
\setcounter{equation}{0}

\resizebox{!}{12cm}{
%{\tiny
\begin{algorithm}[H]
\caption{Dominant Drivers Detection via Network Matrix}
\label{alg:dominant-drivers}
\SetKwInOut{Input}{Input}
\SetKwInOut{Params}{Tuning}
\SetKwInOut{Output}{Output}

\Input{Panel $\{x_{it}\}_{i=1,\dots,N;\ t=1,\dots,T}$}
\Params{%
Selection method $m \in \{\text{LASSO},\ \text{OCMT}\}$; \\
De-factorisation (PCA or CCE, optional); \\
Diagonal-ratio threshold $R \in (0,1]$; connectivity $q \in (0,1]$; \\
Sparsity threshold $\varepsilon>0$ (for link counting);
}
\Output{$\hat N_d$, set of dominant drivers $\widehat{\Gamma}(\hat N_d)$, $\hat B$, $\hat\kappa$.}

\BlankLine
\textbf{(A) Optional factor adjustment}\;
\lIf{ $\hat r > 0$}{
  Estimate $F_{t}\in\mathbb{R}^{\hat r}$ and loadings by PCA or CCE
}

\BlankLine
\textbf{(B) Row-wise sparse regression and residual variances}\;
\For{$i=1$ \KwTo $N$}{
  Define regressor set $\mathcal{P}_i=\{1,\dots,N\}\setminus\{i\}$ and $p=N-1$\;
  \uIf{$m=\text{LASSO}$}{
    Compute rigorous LASSO
    \[
      \hat{\bm b}_i=\arg\min_{\bm b_i\in\mathbb{R}^{p}}
      \frac{1}{T}\sum_{t=1}^T\!\Big(x_{it}-\sum_{j\in\mathcal{P}_i}x_{jt}b_{ij}\Big)^2
      +\lambda_i\sum_{j\in\mathcal{P}_i}\psi_{ij}|b_{ij}|,
    \]
  }
  \uElseIf{$m=\text{OCMT}$}{
    \For{$j\in\mathcal{P}_i$}{
      Run $x_{it}=\alpha_{ij}+\delta_{ij}x_{jt}+e_{ijt}$,\\
      and compute robust $t$-stat / $p$-value $p_{ij}$\;
    }
    Adjust $\{p_{ij}\}$ for multiplicity at level $\alpha$; let $\widehat{\mathcal{S}}_i=\{j:\ \tilde p_{ij}\le \alpha\}$\;
    Refit OLS: $x_{it}=\sum_{j\in\widehat{\mathcal{S}}_i}x_{jt}b_{ij}+u_{it}$; set $b_{ij}=0$ for $j\notin\widehat{\mathcal{S}}_i$ and $b_{ii}=0$.
  }
  Store row $i$ of $\hat B=[\hat b_{ij}]$\;
  Let $\hat\sigma_i^2=\frac{1}{T}\sum_{t=1}^T \hat\epsilon_{it}^2$ be residual variance from the post estimation\;
}
\BlankLine

\textbf{(C) Network matrix construction}\;
Form $\hat D=\mathrm{diag}(1/\hat\sigma_1,\dots,1/\hat\sigma_N)$ and compute
\[
\hat\kappa=\hat D\,(\,I-\hat B\,)\,\hat D.
\]

\BlankLine
\textbf{(D) Candidate filtering (diagonal-ratio and connectivity)}\;
\For{$i=1$ \KwTo $N$}{
  Column norm $c_i=\|\hat\kappa_{(i)}\|$\;
  Diagonal ratio $R_i=|\hat\kappa_{ii}|/c_i$\;
  Connectivity (degree) $d_i=\#\{j\ne i:\ |\hat\kappa_{ji}|>\varepsilon\}$\;
}
Let $\mathcal{I}_R=\{i:\ R_i<R,~ c_i > 1/N\}$ and $\mathcal{I}_d=$ indices with $d_i$ in the top $q$-quantile of $\{d_\ell\}_{\ell=1}^N$\;
Define candidate set $\mathcal{C}=\mathcal{I}_R\cap\mathcal{I}_d$ \;

\BlankLine
\textbf{(E) Ratio rule for the number of dominant drivers}\;
Compute $\{c_i\}_{i\in\mathcal{C}}$, sort in descending order:
$c_{(1)}\ge c_{(2)}\ge\cdots\ge c_{(|\mathcal{C}|)}$\;
Set
\[
\hat N_d=\argmax_{s=1,\dots,|\mathcal{C}|-1}\ \frac{c_{(s)}}{c_{(s+1)}}.
\]
\BlankLine

\textbf{(F) Selection of dominant drivers}\;
Let $\widehat{\Gamma}(\hat N_d)$ be the indices of the top $\hat N_d$ columns in the ordering

\BlankLine
\textbf{Return} $\hat N_d$, $\widehat{\Gamma}(\hat N_d)$, $\hat B$, and $\hat\kappa$.
\end{algorithm}
}

\newpage

\section*{Appendix C: Additional Graphs}
\renewcommand{\theequation}{C.\arabic{equation}}
\renewcommand{\thesection}{C}
\setcounter{equation}{0}

\subsection{Dominant drivers without diagonal restrictions} \label{Appendix: Graphs no restrictions}

\begin{figure}[h]
    \centering
    \includegraphics[width=\linewidth]{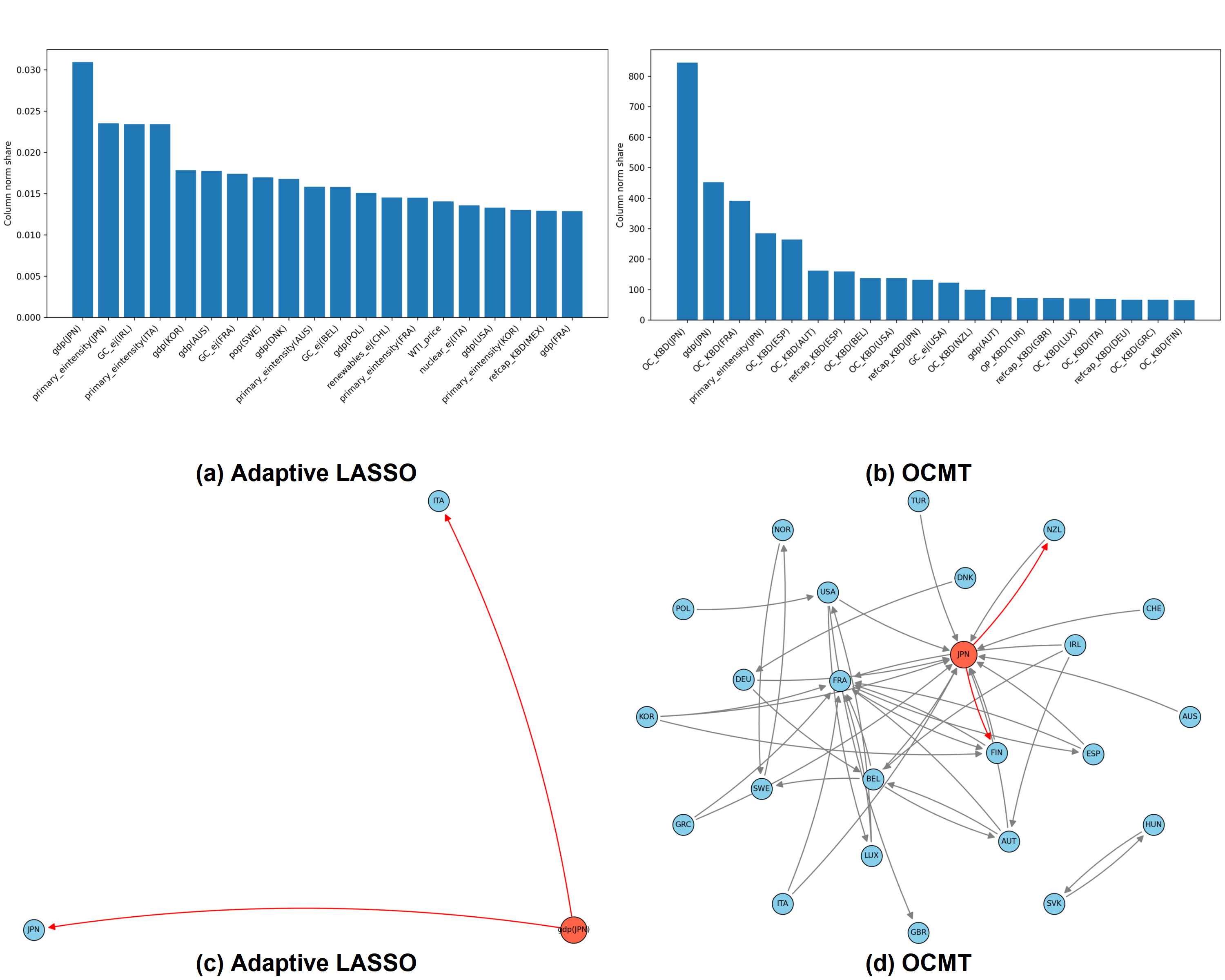}
    \caption{Column norm shares and network graphs - OECD (Unrestricted)}
    \label{fig: norm and network OECD (U)}
\end{figure}

\begin{figure}[h]
    \centering
    \includegraphics[width=\linewidth]{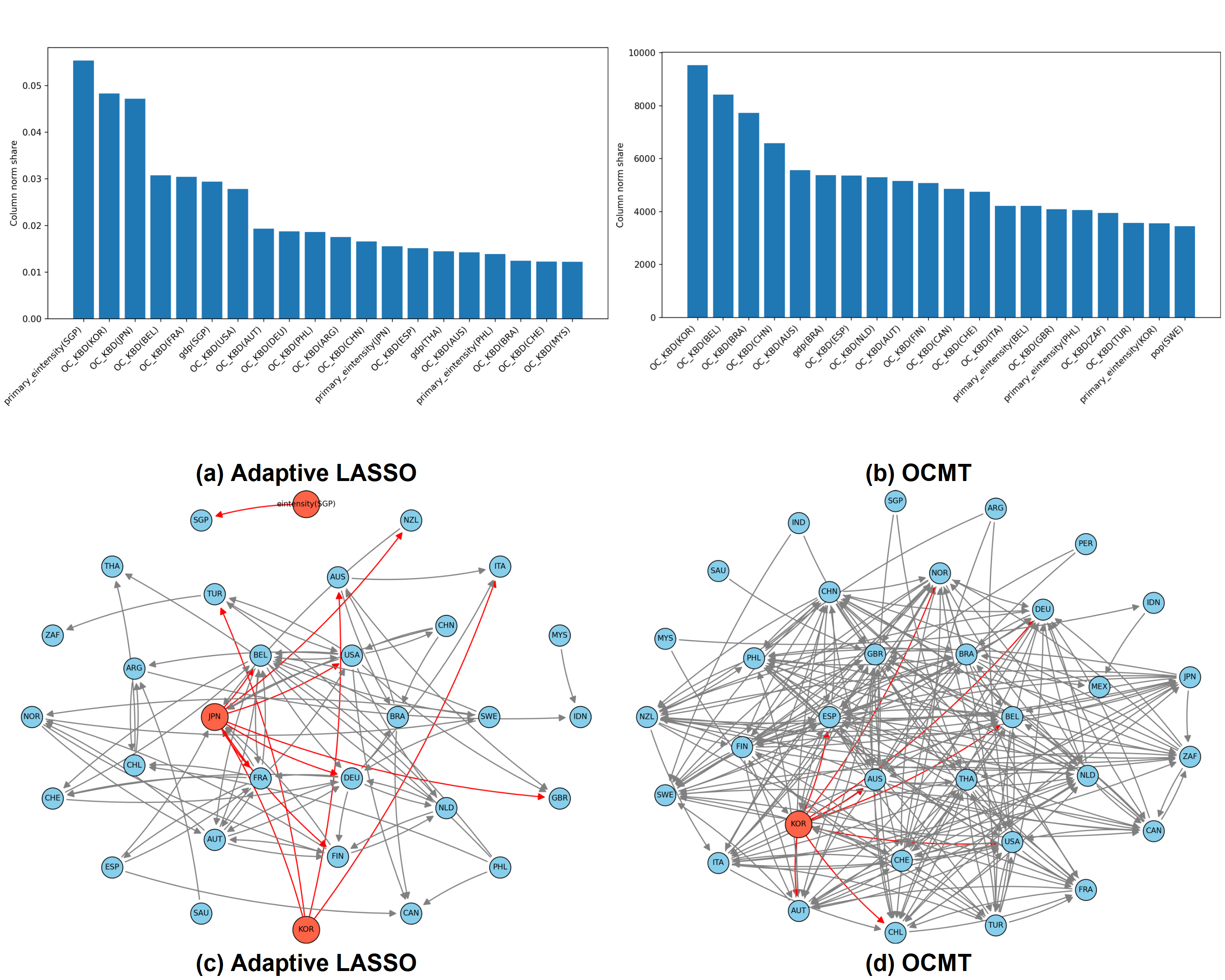}
    \caption{Column norm shares and network graphs - GVAR (Unrestricted)}
    \label{fig: norm and network GVAR (U)}
\end{figure}

\clearpage

\subsection{Controlling for Unobserved Common Factors (CCE)}\label{Appendix: Graphs factors}

\begin{figure}[h]
    \centering
    \includegraphics[width=\linewidth]{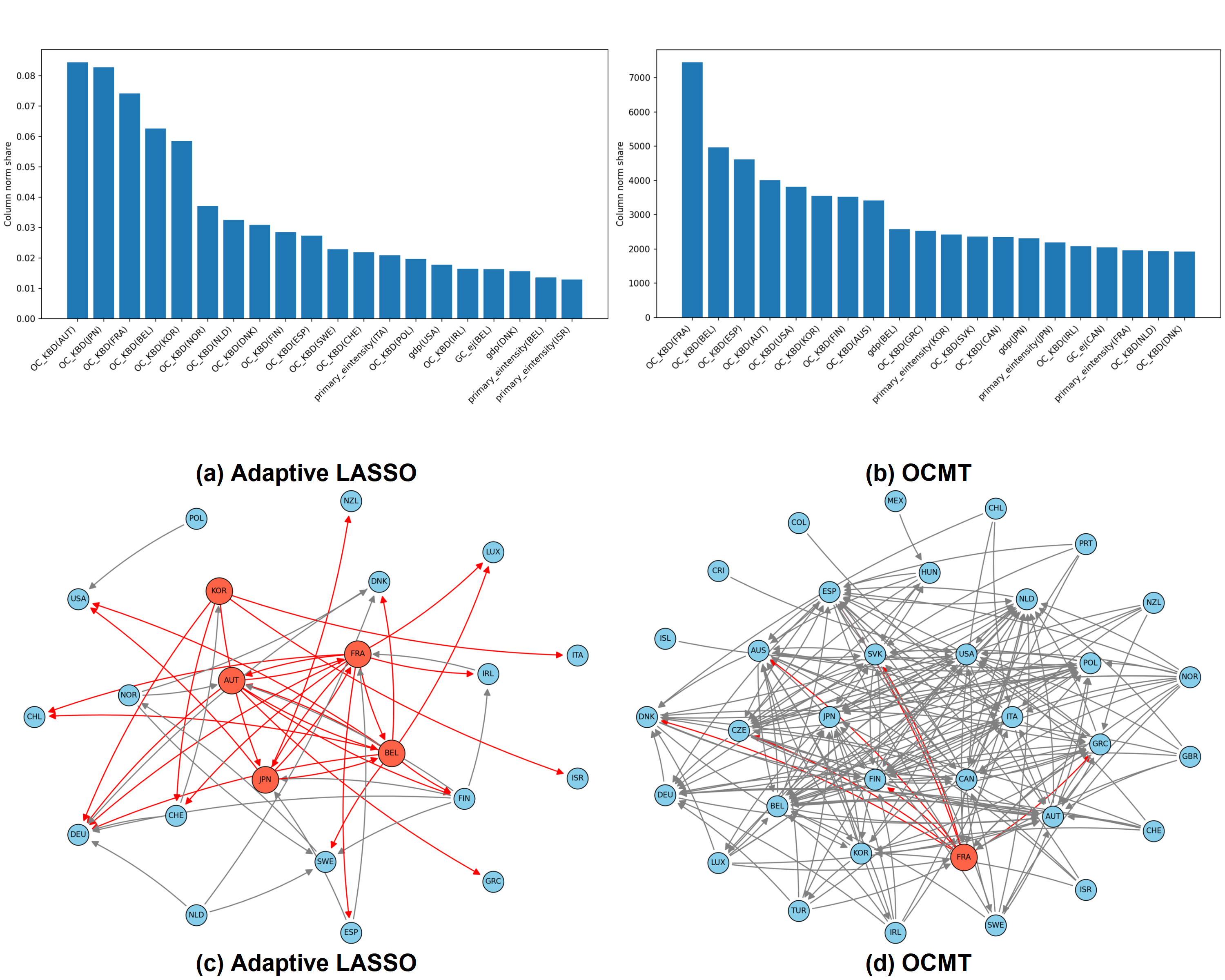}
    \caption{Column norm shares and network graphs - OECD (Unobserved common factors controlled)}
    \label{fig: norm and network OECD PC}
\end{figure}

\begin{figure}[h]
    \centering
    \includegraphics[width=\linewidth]{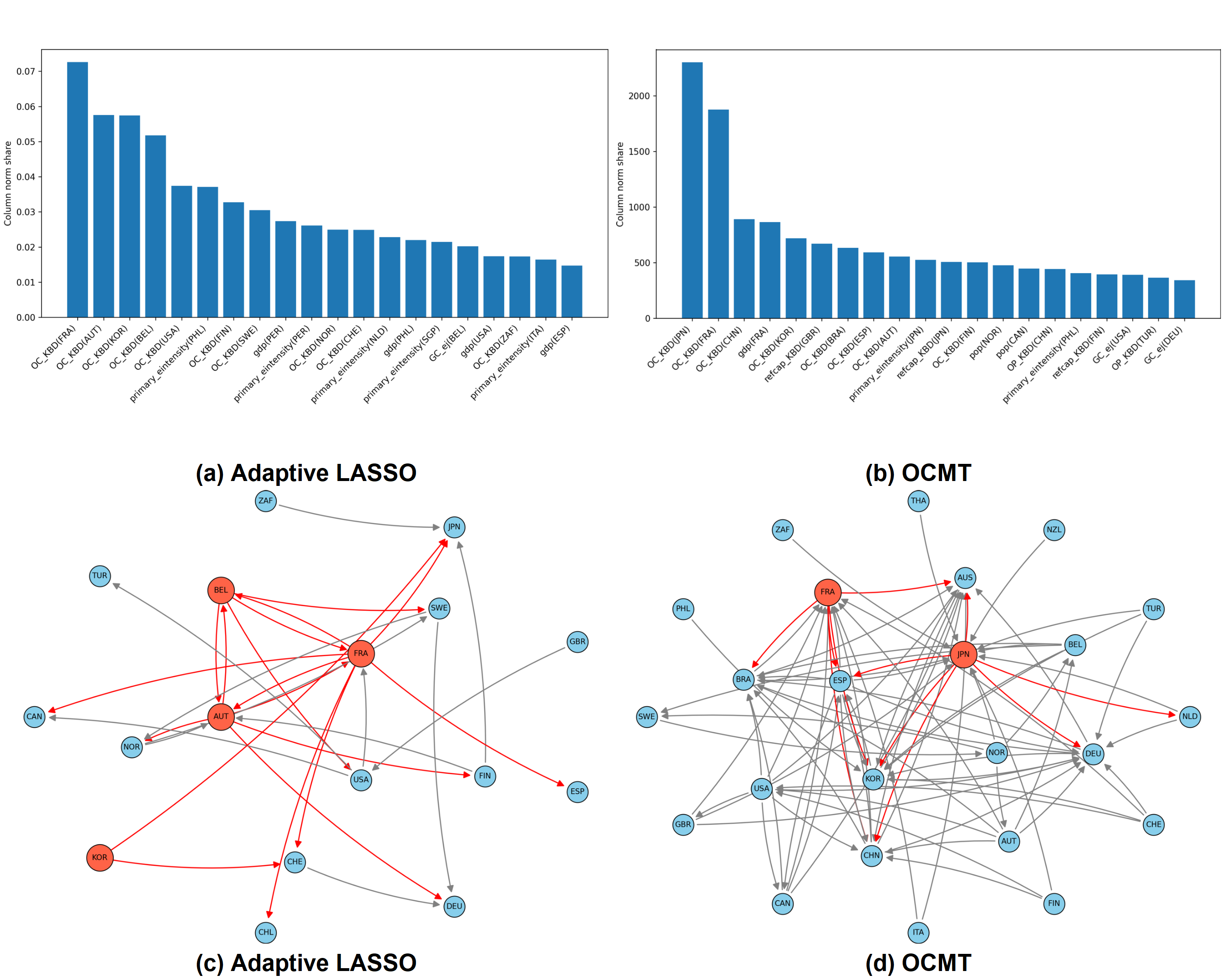}
    \caption{Column norm shares and network graphs - GVAR (Unobserved common factors controlled)}
    \label{fig: norm and network GVAR PC}
\end{figure}

\begin{figure}[htbp]
    \centering
    % First (top) subfigure
    \begin{subfigure}{0.75\textwidth}
        \centering
        \includegraphics[width=\linewidth]{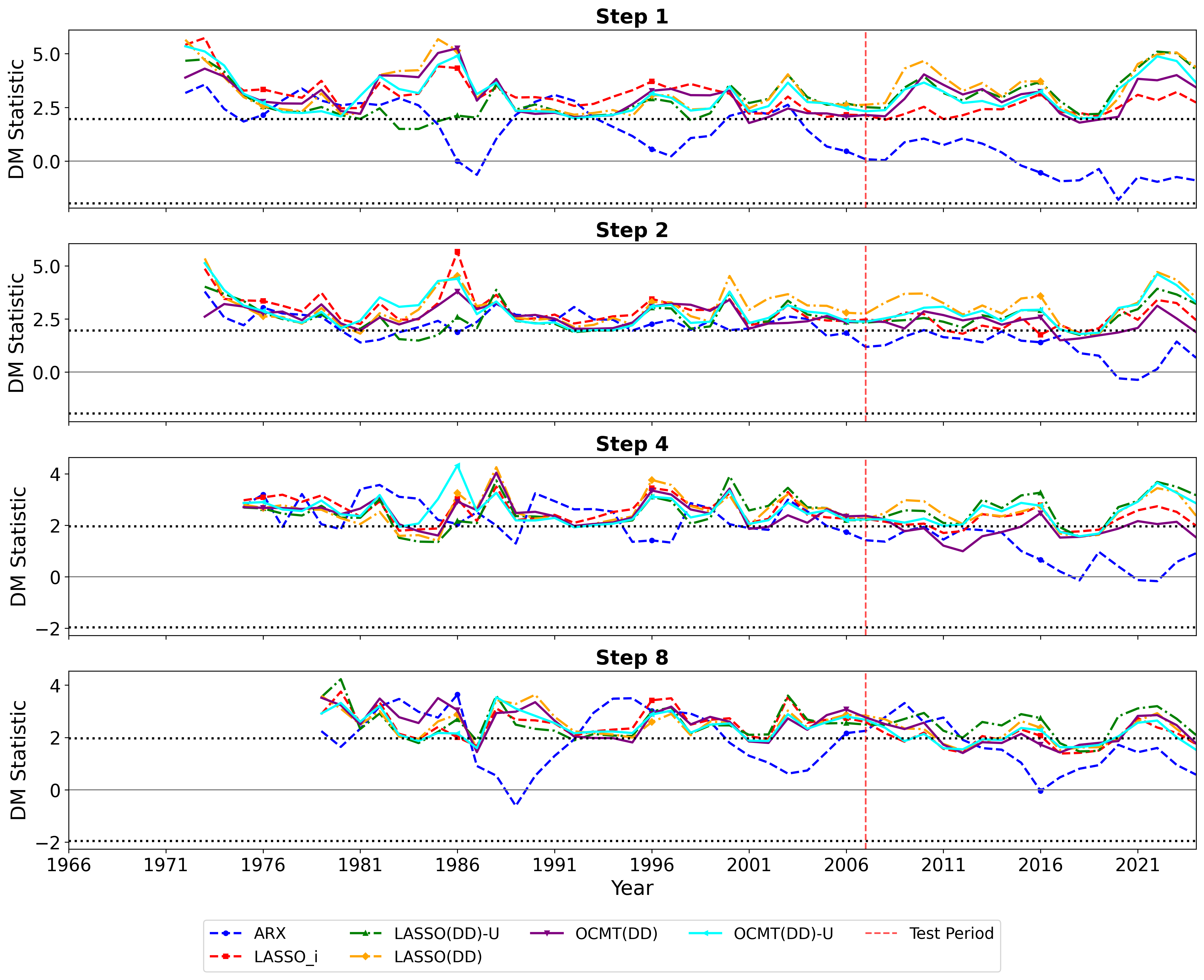}
        \caption{OECD}
        \label{fig:DM forecast OECD-U}
    \end{subfigure}

    \vspace{4mm}

    % Second (bottom) subfigure
    \begin{subfigure}{0.75\textwidth}
        \centering
        \includegraphics[width=\linewidth]{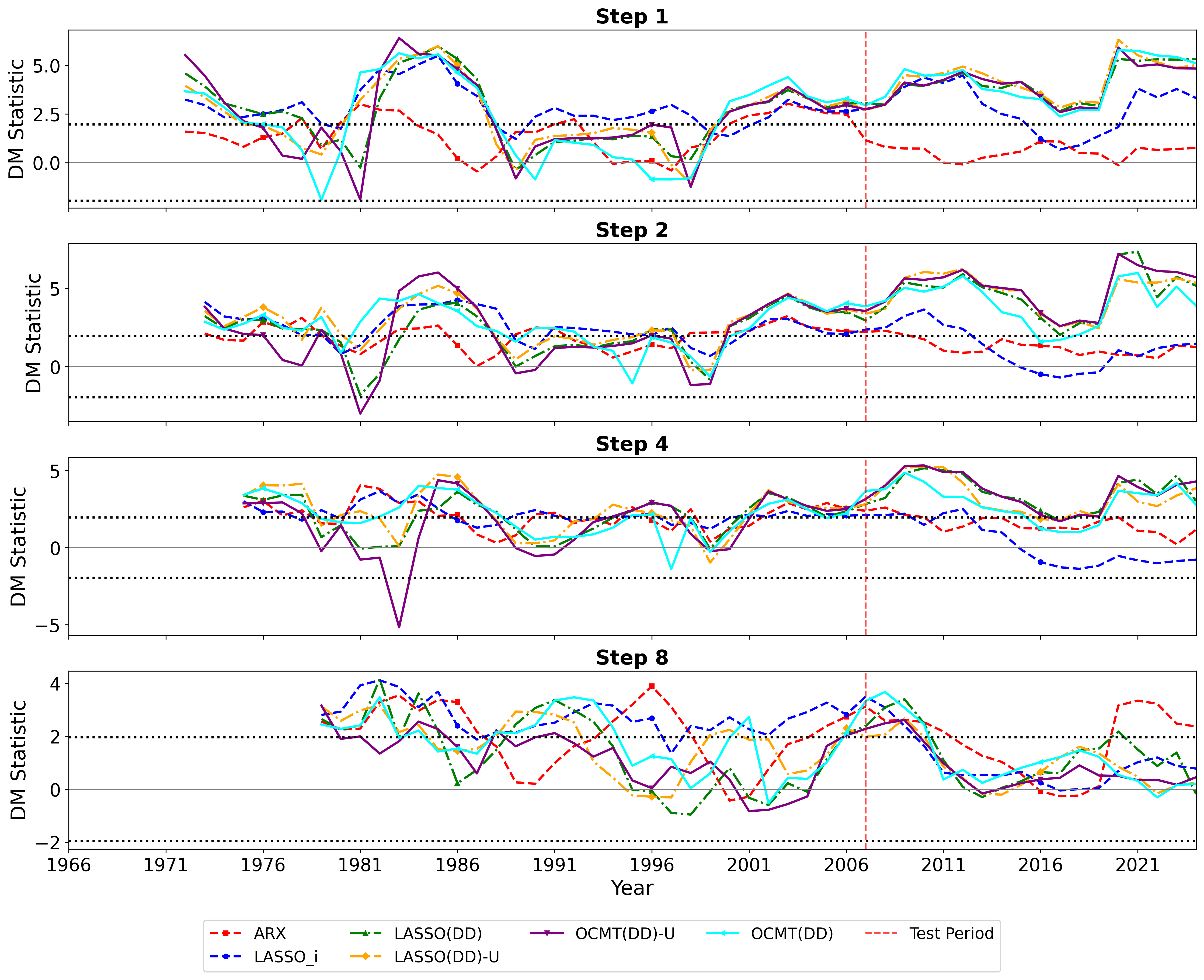}
        \caption{GVAR}
        \label{fig:DM forecast GVAR-U}
    \end{subfigure}

    \caption{Cross-sectional Diebold-Mariano Test statistics – AR benchmark}
    \label{fig: DM forecast Both Unrestricted}
    \vspace{2mm}
    \raggedright
    \footnotesize\textit{Note:} Cross-sectional test statistics for comparisons between competitors and the benchmark with the null of equal forecasts. ARX stands for the AR model with additional regressors, LASSO(DD) and OCMT(DD) are dominant-driver methods, $\text{LASSO}_i$ selects each country's own regressors. The horizontal black dashed lines denote the 5\% significance level, and the vertical dashed line separates the training and testing samples at year 2007.
    \par % end raggedright
\end{figure}

\clearpage

\subsection{SVAR results with dominant drivers excluded from the aggregate demand}
\begin{figure}[H]
    \centering

    \begin{subfigure}{0.75\linewidth}
        \centering
        \includegraphics[width=\linewidth]{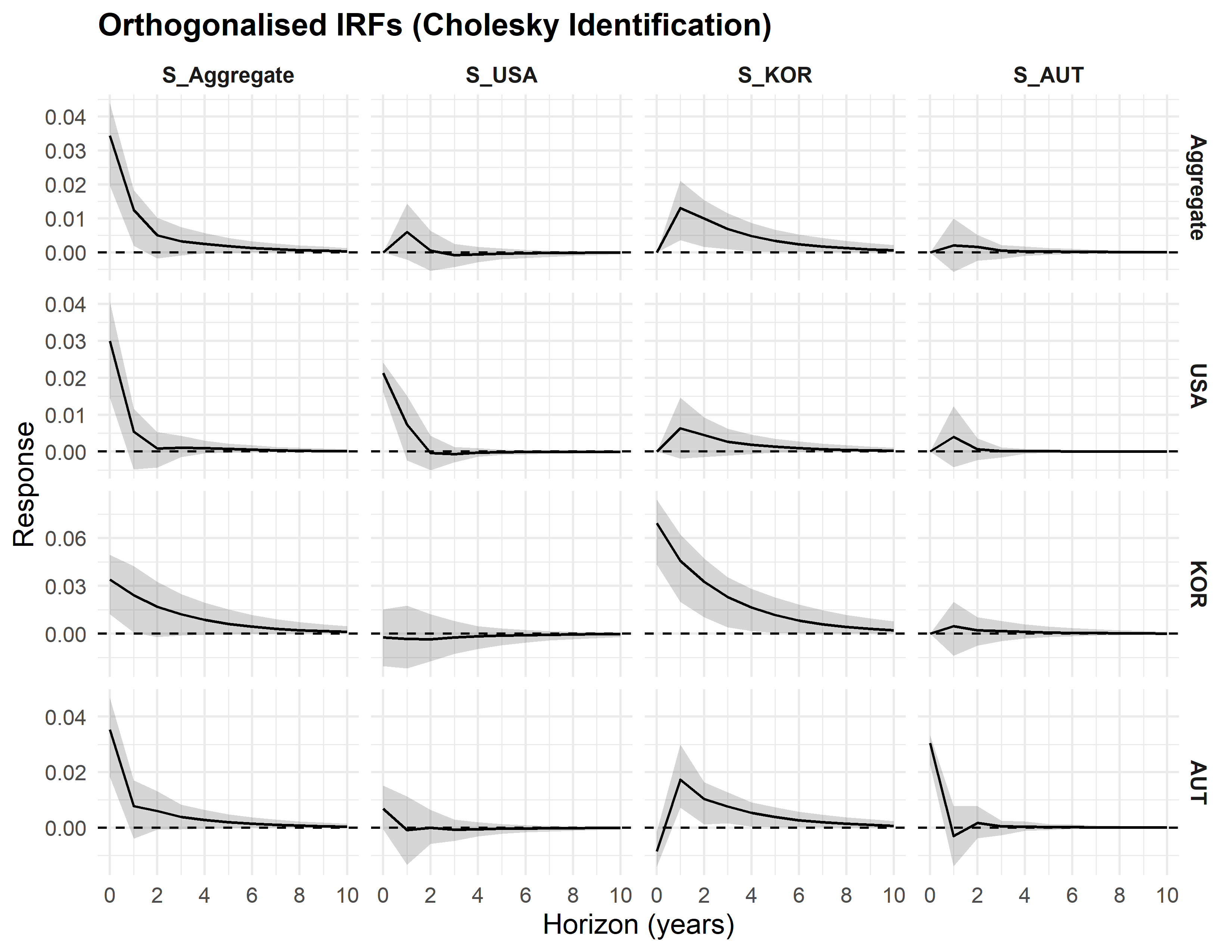}
        \caption{LASSO}
        \label{fig:irf-lasso-oecd-exc}
    \end{subfigure}

    \vspace{0.3em}

    \begin{subfigure}{0.75\linewidth}
        \centering
        \includegraphics[width=\linewidth]{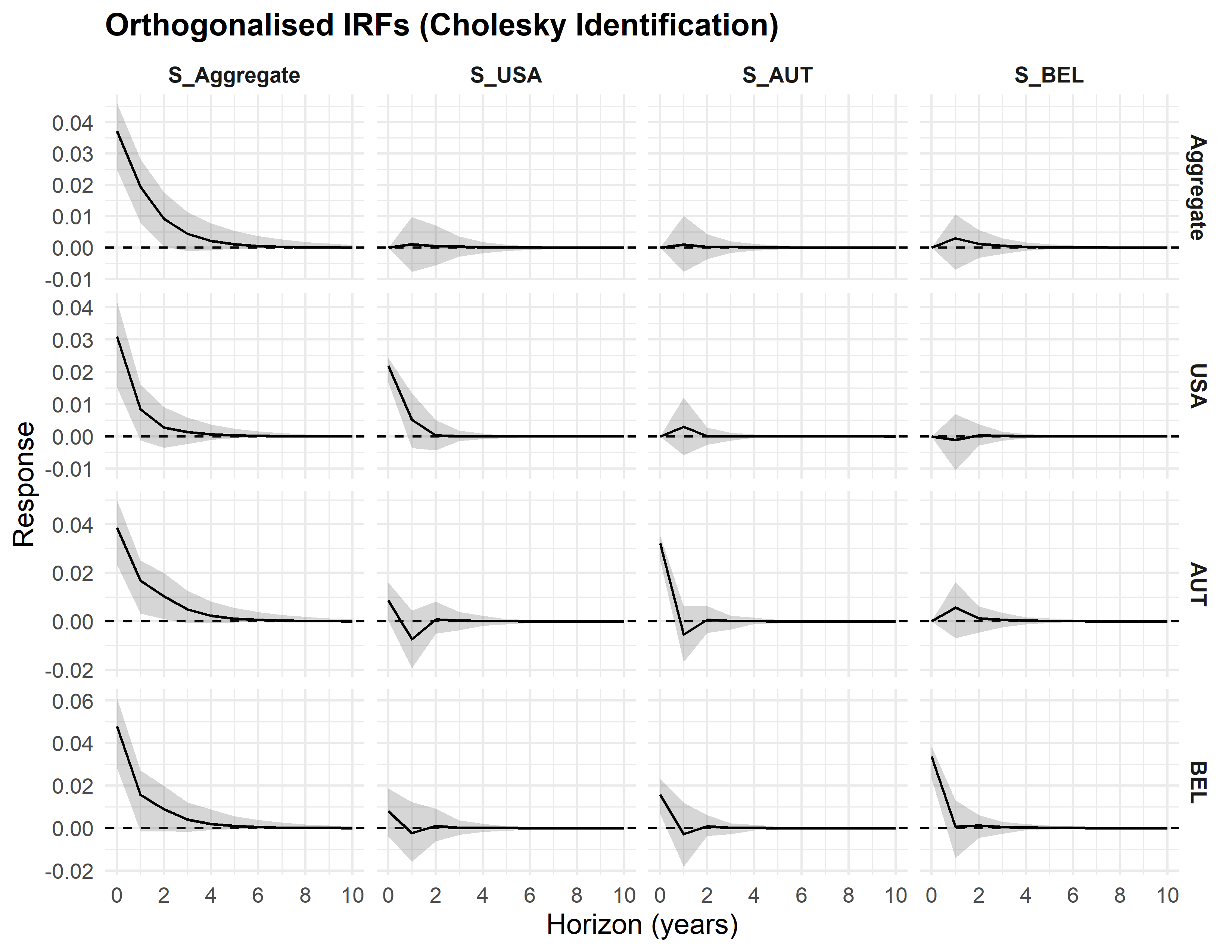}
        \caption{OCMT}
        \label{fig:irf-ocmt-oecd-exc}
    \end{subfigure}

    \caption{Recursive identification (Cholesky) for the OECD sample with dominant drivers excluded from the construction of aggregate series. Columns (S\_$\cdot$) are orthogonalised shocks and rows are responses. The aggregate series is ordered first, followed by dominant-driver countries in the order shown in the panels. 95\% confidence bands with 500 bootstrap replications. VAR lag length (1) selected by AIC.}
    \label{fig:irf-oecd-exc}
\end{figure}

\begin{figure}[htbp]
    \centering

    \begin{subfigure}{0.75\linewidth}
        \centering
        \includegraphics[width=\linewidth]{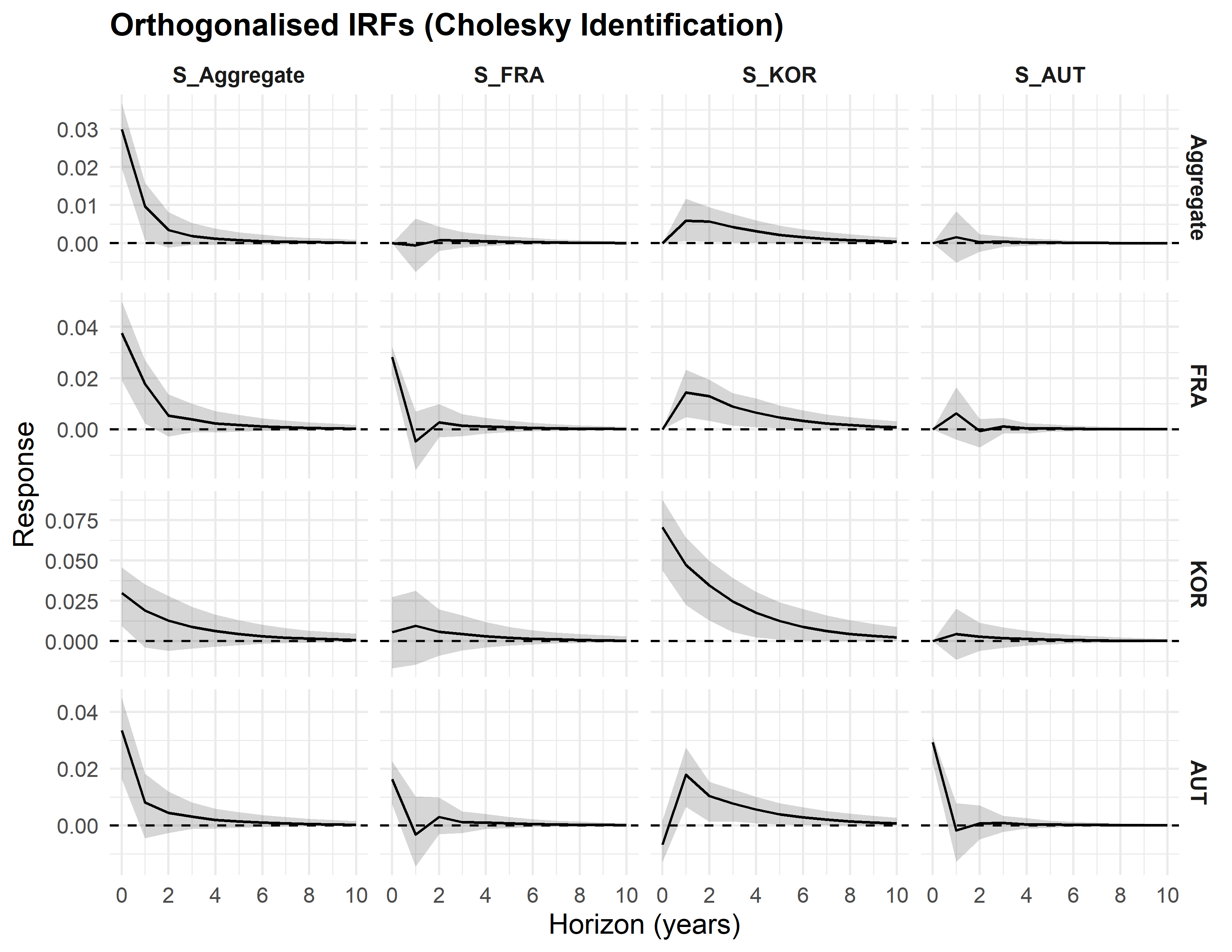}
        \caption{LASSO}
        \label{fig:irf-lasso-GVAR-exc}
    \end{subfigure}

    \vspace{0.3em}

    \begin{subfigure}{0.75\linewidth}
        \centering
        \includegraphics[width=\linewidth]{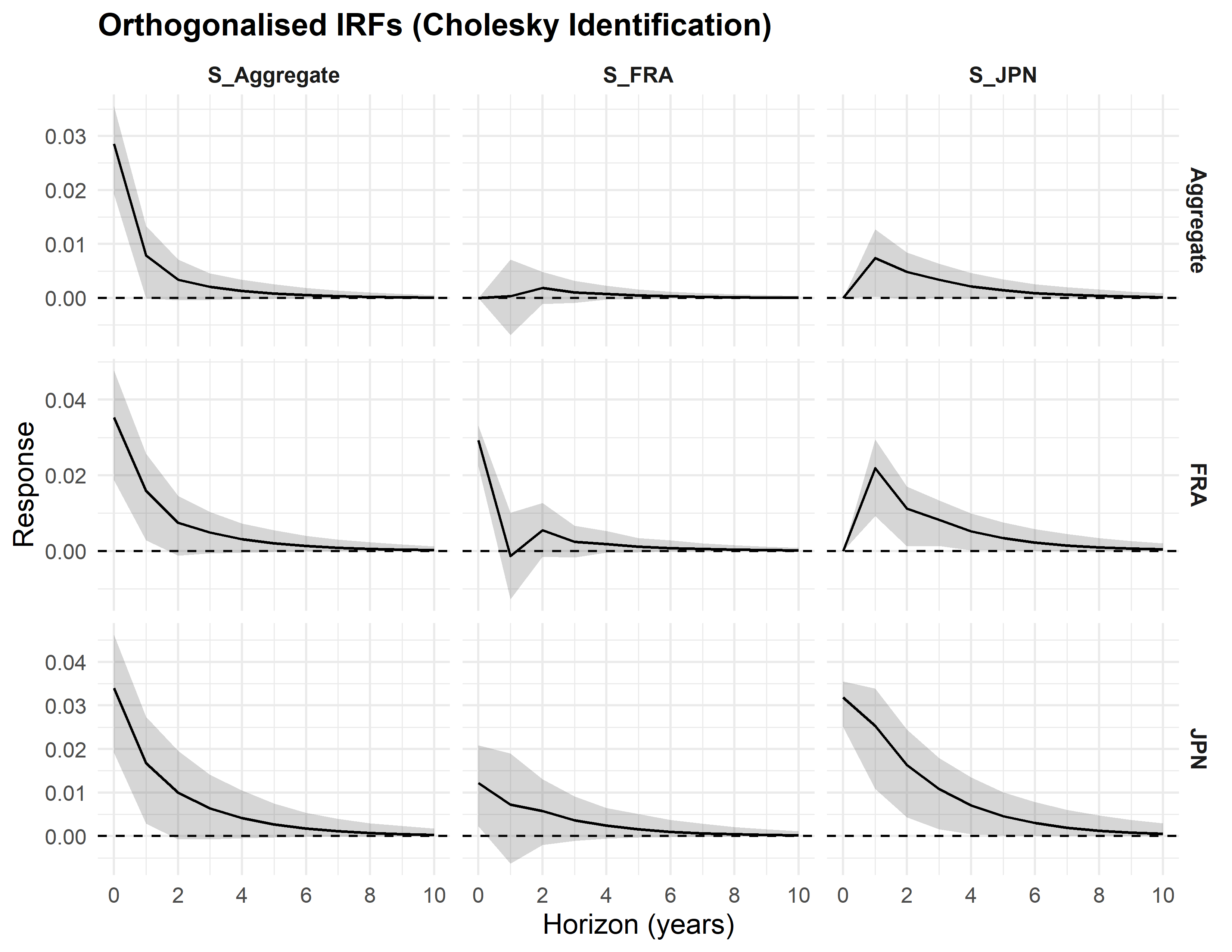}
        \caption{OCMT}
        \label{fig:irf-ocmt-GVAR-exc}
    \end{subfigure}

    \caption{Recursive identification (Cholesky) for the GVAR sample with dominant drivers excluded from the construction of aggregate series. Columns (S\_$\cdot$) are orthogonalised shocks and rows are responses. The aggregate series is ordered first, followed by dominant-driver countries in the order shown in the panels. 95\% confidence bands with 500 bootstrap replications. VAR lag length (1) selected by AIC.}
    \label{fig:irf-GVAR-exc}
\end{figure}

\clearpage
\section*{Appendix D: Additional Tables}
\renewcommand{\theequation}{D.\arabic{equation}}
\renewcommand{\thesection}{D}
\setcounter{equation}{0}

\subsection{Without controlling for unobserved common factors}\label{Appendix: Tables no factors}
\input{Tables/DM_Table_GVAR_NO_F}

\input{Tables/forecast_presentation_table_oecd70_NO_PC}
\input{Tables/forecast_presentation_table_gvar70_NO_PC}

\clearpage

\subsection{Controlling for unobserved common factors (CCE)}\label{Appendix: Tables factors}
\input{Tables/DM_Table_OECD_F}
\input{Tables/DM_Table_GVAR_F}

\input{Tables/forecast_presentation_table_oecd70_PC}
\input{Tables/forecast_presentation_table_gvar70_PC}

%% file: Tables/OECD.tex
\begin{table}[H]
\caption{\scshape OECD Member Countries}
\vspace{0.5em}
\small
\resizebox{0.85\textwidth}{!}{
\begin{tabular}{l l l l l l}\hline\hline
Abbreviation & Country name & Part of EEA & Abbreviation & Country name & Part of EEA\\ \hline
AUS & Australia & No & MEX & Mexico & No\\
AUT & Austria & Yes & NLD & Netherlands & Yes\\
BEL & Belgium & Yes & NOR & Norway & Yes\\
CAN & Canada & No & NZL & New Zealand & No\\
CHE & Switzerland & Yes & POL & Poland & Yes\\
CHL & Chile & No & PRT & Portugal & Yes\\
CZE & Czech Republic & Yes & SVK & Slovak Republic & Yes\\
DEU & Germany & Yes & SWE & Sweden & Yes\\
DNK & Denmark & Yes & TUR & Türkiye & No\\
ESP & Spain & Yes & USA & United States & No\\
FIN & Finland & Yes & GRC & Greece & Yes\\
FRA & France & Yes & HUN & Hungary & Yes\\
GBR & United Kingdom & Yes & IRL & Ireland & Yes\\
ISL & Iceland & Yes & ISR & Israel & No\\
ITA & Italy & Yes & JPN & Japan & No\\
KOR & Korea & No & LUX & Luxembourg & Yes\\ \hline\hline
\end{tabular}}
\\
{\footnotesize \textit{Note:} Colombia, Costa Rica, Estonia, and Slovenia are omitted due to missing data.\\
The United Kingdom left the EEA following Brexit in 2020.}
\label{tab:oecd}
\end{table}

%% file: Tables/GVAR.tex
\begin{table}[H]
\caption{\scshape GVAR Countries}
\vspace{0.5em}
\small
\resizebox{0.85\textwidth}{!}{
\begin{tabular}{l l l l l l}\hline\hline
Abbreviation & Country name & Part of EEA & Abbreviation & Country name & Part of EEA\\ \hline
ARG & Argentina & No & MYS & Malaysia & No\\
AUS & Australia & No & MEX & Mexico & No\\
AUT & Austria & Yes & NLD & Netherlands & Yes\\
BEL & Belgium & Yes & NOR & Norway & Yes\\
BRA & Brazil & No & NZL & New Zealand & No\\
CAN & Canada & No & PER & Peru & No\\
CHE & Switzerland & No & PHL & Philippines & No\\
CHL & Chile & No & SAU & Saudi Arabia & No\\
CHN & China & No & SGP & Singapore & No\\
DEU & Germany & Yes & SWE & Sweden & Yes\\
ESP & Spain & Yes & THA & Thailand & No\\
FIN & Finland & Yes & TUR & Türkiye & No\\
FRA & France & Yes & USA & United States & No\\
GBR & United Kingdom & Yes & ZAF & South Africa & No\\
IDN & Indonesia & No & IND & India & No\\
ITA & Italy & Yes & JPN & Japan & No\\
KOR & Korea & No &  &  & \\ 
\hline\hline
\end{tabular}}
\label{tab:gvar}
\end{table}

%% file: Tables/DM_Table_GVAR_NO_F.tex
{
\begin{table}[htbp]
\centering
\resizebox{0.9\textwidth}{!}{
\begin{tabular}{@{\extracolsep{4pt}}l ccccc @{}}
\hline
\hline
GVAR & ARX & LASSO\(_i\) & LASSO(DD) & OCMT(DD) \\
\hline
\multicolumn{5}{l}{1 Step} \\
\hline
\(1/33 \sum_{i=1}^N I(S_{T,i} \leq 0)\) & 27.27 & 9.09 & 6.06 & 12.12 \\
sig. & 15.15 & 3.03 & 0.00 & 6.06 \\
\(1/33 \sum_{i=1}^N I(S_{T,i} > 0)\) & 72.73 & 90.91 & 93.94 & 87.88 \\
sig. & 60.61 & 87.88 & 90.91 & 84.85 \\
\hline
\multicolumn{5}{l}{2 Step} \\
\hline
\(1/33 \sum_{i=1}^N I(S_{T,i} \leq 0)\) & 33.33 & 9.09 & 3.03 & 3.03 \\
sig. & 24.24 & 6.06 & 0.00 & 3.03 \\
\(1/33 \sum_{i=1}^N I(S_{T,i} > 0)\) & 66.67 & 90.91 & 96.97 & 96.97 \\
sig. & 51.52 & 75.76 & 84.85 & 84.85 \\
\hline
\multicolumn{5}{l}{4 Step} \\
\hline
\(1/33 \sum_{i=1}^N I(S_{T,i} \leq 0)\) & 30.30 & 30.30 & 9.09 & 12.12 \\
sig. & 18.18 & 12.12 & 0.00 & 0.00 \\
\(1/33 \sum_{i=1}^N I(S_{T,i} > 0)\) & 69.70 & 69.70 & 90.91 & 87.88 \\
sig. & 48.48 & 48.48 & 75.76 & 69.70 \\
\hline
\multicolumn{5}{l}{8 Step} \\
\hline
\(1/33 \sum_{i=1}^N I(S_{T,i} \leq 0)\) & 12.12 & 12.12 & 15.15 & 15.15 \\
sig. & 3.03 & 6.06 & 6.06 & 6.06 \\
\(1/33 \sum_{i=1}^N I(S_{T,i} > 0)\) & 87.88 & 87.88 & 84.85 & 84.85 \\
sig. & 63.64 & 63.64 & 39.39 & 30.30 \\
\hline
\hline
\end{tabular}}
\caption{Test of equal forecast accuracy against a simple AR. \(S_{T,i}\) is the share of countries with a positive or negative test statistic from the Diebold-Mariano test \citep{Diebold1995}. ``Sig.'' indicates the percentage share of cases where \(S_{T,i}\) are significantly positive or negative at the 5\% level. A positive \(S_{T,i}\) implies that forecasts based on dominant drivers are more accurate. ARX is the AR model with forced fundamentals included as regressors, $\text{LASSO}_i$ is the country specific variable selection method, LASSO(DD) and OCMT(DD) are dominant-driver methods with restricted diagonal ratio.}
\label{table:DM_GVAR no factors}
\end{table}
}

%% file: Tables/forecast_presentation_table_oecd70_NO_PC.tex
\begin{table}[htbp]
\centering
\caption{Forecast evaluation -- OECD (Unobserved common factors uncontrolled)}
\label{tab:forecast_oecd_NO_PC}
\begin{tabular}{lccccc}
\toprule
Model & RMSE & MAE & RMSE ratio & MAE ratio & MCS rate \\
\midrule
\multicolumn{6}{l}{\textbf{Horizon 1}} \\
AR & 0.0686 & 0.0503 & 100.00\% & 100.00\% & 93.33\% \\
ARX & 0.0826 & 0.0638 & 125.58\% & 126.75\% & 73.33\% \\
LASSO\_i & 0.0734 & 0.0555 & 107.22\% & 110.29\% & 66.67\% \\
LASSO(DD) & 0.0667 & 0.0483 & 97.22\% & 96.08\% & 93.33\% \\
LASSO(DD)-U & 0.0645 & 0.0476 & 94.23\% & 94.62\% & 90.00\% \\

OCMT(DD) & 0.0699 & 0.0508 & 101.33\% & 100.91\% & 86.67\% \\
OCMT(DD)-U & 0.0663 & 0.0486 & 96.61\% & 96.49\% & 73.33\% \\
\midrule
\multicolumn{6}{l}{\textbf{Horizon 2}} \\
AR & 0.0673 & 0.0491 & 100.00\% & 100.00\% & 93.33\% \\
ARX & 0.0756 & 0.0565 & 111.60\% & 115.02\% & 70.00\% \\
LASSO\_i & 0.0735 & 0.0550 & 114.84\% & 112.03\% & 90.00\% \\
LASSO(DD) & 0.0677 & 0.0497 & 100.76\% & 101.26\% & 93.33\% \\
LASSO(DD)-U & 0.0656 & 0.0483 & 97.49\% & 98.27\% & 90.00\% \\

OCMT(DD) & 0.0712 & 0.0521 & 105.28\% & 106.10\% & 80.00\% \\
OCMT(DD)-U & 0.0657 & 0.0489 & 97.54\% & 99.67\% & 90.00\% \\
\midrule
\multicolumn{6}{l}{\textbf{Horizon 4}} \\
AR & 0.0663 & 0.0483 & 100.00\% & 100.00\% & 90.00\% \\
ARX & 0.0756 & 0.0579 & 113.20\% & 119.89\% & 66.67\% \\
LASSO\_i & 0.0695 & 0.0520 & 104.41\% & 107.79\% & 80.00\% \\
LASSO(DD) & 0.0664 & 0.0486 & 99.76\% & 100.75\% & 90.00\% \\
LASSO(DD)-U & 0.0642 & 0.0475 & 96.70\% & 98.46\% & 90.00\% \\

OCMT(DD) & 0.0682 & 0.0506 & 102.32\% & 104.74\% & 73.33\% \\
OCMT(DD)-U & 0.0641 & 0.0477 & 96.46\% & 98.88\% & 93.33\% \\
\midrule
\multicolumn{6}{l}{\textbf{Horizon 8}} \\
AR & 0.0646 & 0.0467 & 100.00\% & 100.00\% & 96.67\% \\
ARX & 0.0712 & 0.0533 & 112.80\% & 114.27\% & 80.00\% \\
LASSO\_i & 0.0671 & 0.0498 & 103.50\% & 106.70\% & 86.67\% \\
LASSO(DD) & 0.0674 & 0.0499 & 104.06\% & 106.79\% & 80.00\% \\
LASSO(DD)-U & 0.0658 & 0.0483 & 101.37\% & 103.53\% & 83.33\% \\

OCMT(DD) & 0.0677 & 0.0498 & 104.37\% & 106.71\% & 83.33\% \\
OCMT(DD)-U & 0.0660 & 0.0481 & 101.68\% & 103.06\% & 90.00\% \\
\midrule
\bottomrule
\end{tabular}

\vspace{0.5cm}
{\footnotesize
\begin{minipage}{\linewidth}
\raggedright
\textit{Note:} LASSO(DD)-U and OCMT(DD)-U indicate dominant drivers sets selected without diagonal ratio restriction. Both RMSE and MAE ratios are compared against the AR model. MCS rate reports the fraction of countries in which the model belongs to the 10\% Model Confidence Set \citep{hansen2011model} computed on squared forecast errors using a moving-block bootstrap (block length 3, 2000 replications) and the range statistic based on the maximum absolute studentised pairwise mean loss differential across models. The procedure iteratively drops the model performs the worst in the comparison. A higher MCS rate implies the model is more competitive in terms of squared loss. 
\end{minipage}
}

\end{table}

%% file: Tables/forecast_presentation_table_gvar70_NO_PC.tex
{\small
\begin{table}[htbp]
\centering
\caption{Forecast evaluation -- GVAR (Unobserved common factors uncontrolled)}
\label{tab:forecast_gvar_NO_PC}
\begin{tabular}{lccccc}
\toprule
Model & RMSE & MAE & RMSE ratio & MAE ratio & MCS rate \\
\midrule
\multicolumn{6}{l}{\textbf{Horizon 1}} \\
AR & 0.0612 & 0.0457 & 100.00\% & 100.00\% & 63.64\% \\
ARX & 0.0739 & 0.0592 & 127.26\% & 129.57\% & 72.73\% \\
LASSO\_i & 0.0687 & 0.0523 & 113.69\% & 114.62\% & 63.64\% \\
LASSO(DD) & 0.0589 & 0.0434 & 97.16\% & 95.11\% & 87.88\% \\
LASSO(DD)-U & 0.0583 & 0.0430 & 95.69\% & 94.18\% & 87.88\% \\
OCMT(DD) & 0.0594 & 0.0442 & 97.55\% & 96.91\% & 84.85\% \\
OCMT(DD)-U & 0.0559 & 0.0413 & 92.16\% & 90.44\% & 96.97\% \\
\midrule
\multicolumn{6}{l}{\textbf{Horizon 2}} \\
AR & 0.0604 & 0.0448 & 100.00\% & 100.00\% & 75.76\% \\
ARX & 0.0702 & 0.0540 & 118.14\% & 120.37\% & 63.64\% \\
LASSO\_i & 0.0675 & 0.0522 & 116.81\% & 116.31\% & 63.64\% \\
LASSO(DD) & 0.0573 & 0.0427 & 95.87\% & 95.25\% & 87.88\% \\
LASSO(DD)-U & 0.0571 & 0.0427 & 95.32\% & 95.25\% & 100.00\% \\
OCMT(DD) & 0.0597 & 0.0449 & 98.96\% & 100.13\% & 87.88\% \\
OCMT(DD)-U & 0.0563 & 0.0417 & 94.26\% & 93.00\% & 96.97\% \\
\midrule
\multicolumn{6}{l}{\textbf{Horizon 4}} \\
AR & 0.0591 & 0.0437 & 100.00\% & 100.00\% & 72.73\% \\
ARX & 0.0964 & 0.0774 & 261.54\% & 176.96\% & 48.48\% \\
LASSO\_i & 0.0745 & 0.0568 & 140.56\% & 129.94\% & 66.67\% \\
LASSO(DD) & 0.0582 & 0.0431 & 98.68\% & 98.50\% & 96.97\% \\
LASSO(DD)-U & 0.0587 & 0.0440 & 99.53\% & 100.61\% & 81.82\% \\
OCMT(DD) & 0.0597 & 0.0455 & 100.95\% & 103.96\% & 81.82\% \\
OCMT(DD)-U & 0.0576 & 0.0422 & 97.84\% & 96.60\% & 90.91\% \\
\midrule
\multicolumn{6}{l}{\textbf{Horizon 8}} \\
AR & 0.0577 & 0.0426 & 100.00\% & 100.00\% & 84.85\% \\
ARX & 0.0637 & 0.0499 & 114.82\% & 117.18\% & 81.82\% \\
LASSO\_i & 0.0677 & 0.0507 & 128.07\% & 118.95\% & 81.82\% \\
LASSO(DD) & 0.0594 & 0.0443 & 102.80\% & 104.03\% & 96.97\% \\
LASSO(DD)-U & 0.0588 & 0.0438 & 102.15\% & 102.79\% & 87.88\% \\
OCMT(DD) & 0.0588 & 0.0434 & 101.99\% & 101.97\% & 93.94\% \\
OCMT(DD)-U & 0.0581 & 0.0431 & 100.76\% & 101.24\% & 84.85\% \\
\midrule
\bottomrule
\end{tabular}

\vspace{0.5cm}
{\footnotesize
\begin{minipage}{\linewidth}
\raggedright
\textit{Note:} LASSO(DD)-U and OCMT(DD)-U indicate dominant drivers sets selected without diagonal ratio restriction. Both RMSE and MAE ratios are compared against the AR model. MCS rate reports the fraction of countries in which the model belongs to the 10\% Model Confidence Set \citep{hansen2011model} computed on squared forecast errors using a moving-block bootstrap (block length 3, 2000 replications) and the range statistic based on the maximum absolute studentised pairwise mean loss differential across models. The procedure iteratively drops the model performs the worst in the comparison. A higher MCS rate implies the model is more competitive in terms of squared loss. 
\end{minipage}
}
\end{table}
}

%% file: Tables/DM_Table_OECD_F.tex
\begin{table}[htbp]
\tiny
\centering
\resizebox{\textwidth}{!}{
\begin{tabular}{@{\extracolsep{4pt}}l cccccc @{}}\hline
\hline
OECD & ARX & LASSO\(_i\) & LASSO(DD) & OCMT(DD) & LASSO(DD)-U & OCMT(DD)-U \\
\hline
\multicolumn{7}{l}{1 Step} \\
\hline
\(1/34 \sum_{i=1}^N I(S_{T,i} \leq 0)\) & 30.00 & 6.67 & 0.00 & 6.67 & 0.00 & 6.67 \\
sig. & 20.00 & 0.00 & 0.00 & 0.00 & 0.00 & 3.33 \\
\(1/34 \sum_{i=1}^N I(S_{T,i} > 0)\) & 70.00 & 93.33 & 100.00 & 93.33 & 100.00 & 93.33 \\
sig. & 63.33 & 83.33 & 93.33 & 76.67 & 90.00 & 80.00 \\
\hline
\multicolumn{7}{l}{2 Step} \\
\hline
\(1/34 \sum_{i=1}^N I(S_{T,i} \leq 0)\) & 40.00 & 16.67 & 3.33 & 6.67 & 3.33 & 13.33 \\
sig. & 23.33 & 3.33 & 0.00 & 3.33 & 0.00 & 6.67 \\
\(1/34 \sum_{i=1}^N I(S_{T,i} > 0)\) & 60.00 & 83.33 & 96.67 & 93.33 & 96.67 & 86.67 \\
sig. & 60.00 & 80.00 & 93.33 & 73.33 & 83.33 & 73.33 \\
\hline
\multicolumn{7}{l}{4 Step} \\
\hline
\(1/34 \sum_{i=1}^N I(S_{T,i} \leq 0)\) & 40.00 & 13.33 & 13.33 & 13.33 & 13.33 & 13.33 \\
sig. & 16.67 & 6.67 & 3.33 & 6.67 & 6.67 & 3.33 \\
\(1/34 \sum_{i=1}^N I(S_{T,i} > 0)\) & 60.00 & 86.67 & 86.67 & 86.67 & 86.67 & 86.67 \\
sig. & 46.67 & 70.00 & 80.00 & 73.33 & 83.33 & 73.33 \\
\hline
\multicolumn{7}{l}{8 Step} \\
\hline
\(1/34 \sum_{i=1}^N I(S_{T,i} \leq 0)\) & 30.00 & 23.33 & 16.67 & 16.67 & 16.67 & 26.67 \\
sig. & 6.67 & 10.00 & 6.67 & 6.67 & 6.67 & 10.00 \\
\(1/34 \sum_{i=1}^N I(S_{T,i} > 0)\) & 70.00 & 76.67 & 83.33 & 83.33 & 83.33 & 73.33 \\
sig. & 56.67 & 66.67 & 70.00 & 63.33 & 66.67 & 60.00 \\
\hline\hline
\end{tabular}}
\captionsetup{width=\textwidth}
\caption{Test of equal forecast accuracy against a simple AR with factors controlled for OECD countries. \(S_{T,i}\) is the share of countries with a positive or negative test statistic from the Diebold-Mariano test \citep{Diebold1995}. ``Sig.'' indicates how often \(S_{T,i}\) are significantly positive or negative at a 5\% level. A positive \(S_{T,i}\) implies that forecasts based on dominant drivers are more accurate. ARX is the AR model with forced fundamentals included as regressors, $\text{LASSO}_i$ is the country specific variable selection method, LASSO(DD) and OCMT(DD) are dominant-driver methods with restricted diagonal ratio, while LASSO(DD)-U and OCMT(DD)-U indicate dominant-driver methods without such restriction.}
\label{table: DM OECD Factors}
\end{table}

%% file: Tables/DM_Table_GVAR_F.tex
\begin{table}[htbp]
\tiny
\centering
\resizebox{\textwidth}{!}{
\begin{tabular}{@{\extracolsep{4pt}}l cccccc @{}}\hline
\hline
GVAR & ARX & LASSO\(_i\) & LASSO(DD) & OCMT(DD) & LASSO(DD)-U & OCMT(DD)-U \\
\hline
\multicolumn{7}{l}{1 Step} \\
\hline
\(1/33 \sum_{i=1}^N I(S_{T,i} \leq 0)\) & 27.27 & 15.15 & 3.03 & 12.12 & 3.03 & 3.03 \\
sig. & 15.15 & 9.09 & 0.00 & 6.06 & 0.00 & 0.00 \\
\(1/33 \sum_{i=1}^N I(S_{T,i} > 0)\) & 72.73 & 84.85 & 96.97 & 87.88 & 96.97 & 96.97 \\
sig. & 60.61 & 78.79 & 90.91 & 84.85 & 87.88 & 90.91 \\
\hline
\multicolumn{7}{l}{2 Step} \\
\hline
\(1/33 \sum_{i=1}^N I(S_{T,i} \leq 0)\) & 33.33 & 15.15 & 3.03 & 3.03 & 6.06 & 12.12 \\
sig. & 24.24 & 6.06 & 0.00 & 3.03 & 0.00 & 0.00 \\
\(1/33 \sum_{i=1}^N I(S_{T,i} > 0)\) & 66.67 & 84.85 & 96.97 & 96.97 & 93.94 & 87.88 \\
sig. & 51.52 & 75.76 & 87.88 & 84.85 & 90.91 & 87.88 \\
\hline
\multicolumn{7}{l}{4 Step} \\
\hline
\(1/33 \sum_{i=1}^N I(S_{T,i} \leq 0)\) & 30.30 & 15.15 & 3.03 & 12.12 & 12.12 & 12.12 \\
sig. & 18.18 & 3.03 & 0.00 & 0.00 & 0.00 & 3.03 \\
\(1/33 \sum_{i=1}^N I(S_{T,i} > 0)\) & 69.70 & 84.85 & 96.97 & 87.88 & 87.88 & 87.88 \\
sig. & 48.48 & 48.48 & 72.73 & 69.70 & 72.73 & 75.76 \\
\hline
\multicolumn{7}{l}{8 Step} \\
\hline
\(1/33 \sum_{i=1}^N I(S_{T,i} \leq 0)\) & 12.12 & 12.12 & 15.15 & 15.15 & 36.36 & 45.45 \\
sig. & 3.03 & 0.00 & 6.06 & 6.06 & 15.15 & 18.18 \\
\(1/33 \sum_{i=1}^N I(S_{T,i} > 0)\) & 87.88 & 87.88 & 84.85 & 84.85 & 63.64 & 54.55 \\
sig. & 63.64 & 60.61 & 51.52 & 30.30 & 36.36 & 36.36 \\
\hline\hline
\end{tabular}}
\captionsetup{width=\textwidth}
\caption{Test of equal forecast accuracy against a simple AR with factors controlled for GVAR countries. \(S_{T,i}\) is the share of regions with a positive or negative test statistic from the Diebold-Mariano test \citep{Diebold1995}. ``Sig.'' indicates how often \(S_{T,i}\) are significantly positive or negative at a 5\% level. A positive \(S_{T,i}\) implies that forecasts based on dominant drivers are more accurate. ARX is the AR model with forced fundamentals included as regressors, $\text{LASSO}_i$ is the country specific variable selection method, LASSO(DD) and OCMT(DD) are dominant-driver methods with restricted diagonal ratio, while LASSO(DD)-U and OCMT(DD)-U indicate dominant-driver methods without such restriction.}
\label{table: DM GVAR Factors}
\end{table}

%% file: Tables/forecast_presentation_table_oecd70_PC.tex
\begin{table}[htbp]
\centering
\caption{Forecast evaluation -- OECD (Unobserved common factors controlled)}
\label{tab:forecast_oecd_PC}
\begin{tabular}{lccccc}
\toprule
Model & RMSE & MAE & RMSE ratio & MAE ratio & MCS rate \\
\midrule
\multicolumn{6}{l}{\textbf{Horizon 1}} \\
AR & 0.0686 & 0.0503 & 100.00\% & 100.00\% & 96.67\% \\
ARX & 0.0826 & 0.0638 & 125.58\% & 126.75\% & 70.00\% \\
LASSO\_i & 0.0745 & 0.0564 & 108.29\% & 111.99\% & 73.33\% \\
LASSO(DD) & 0.0667 & 0.0487 & 97.37\% & 96.80\% & 86.67\% \\
LASSO(DD)-U & 0.0638 & 0.0469 & 93.62\% & 93.13\% & 96.67\% \\

OCMT(DD) & 0.0679 & 0.0496 & 98.81\% & 98.55\% & 73.33\% \\
OCMT(DD)-U & 0.0645 & 0.0476 & 94.23\% & 94.62\% & 93.33\% \\
\midrule
\multicolumn{6}{l}{\textbf{Horizon 2}} \\
AR & 0.0673 & 0.0491 & 100.00\% & 100.00\% & 93.33\% \\
ARX & 0.0756 & 0.0565 & 111.60\% & 115.02\% & 70.00\% \\
LASSO\_i & 0.0732 & 0.0545 & 114.59\% & 111.04\% & 86.67\% \\
LASSO(DD) & 0.0668 & 0.0498 & 99.85\% & 101.37\% & 90.00\% \\
LASSO(DD)-U & 0.0656 & 0.0483 & 97.69\% & 98.42\% & 86.67\% \\

OCMT(DD) & 0.0673 & 0.0494 & 99.96\% & 100.66\% & 76.67\% \\
OCMT(DD)-U & 0.0656 & 0.0483 & 97.49\% & 98.27\% & 93.33\% \\
\midrule
\multicolumn{6}{l}{\textbf{Horizon 4}} \\
AR & 0.0663 & 0.0483 & 100.00\% & 100.00\% & 86.67\% \\
ARX & 0.0756 & 0.0579 & 113.20\% & 119.89\% & 66.67\% \\
LASSO\_i & 0.0700 & 0.0524 & 105.29\% & 108.52\% & 73.33\% \\
LASSO(DD) & 0.0661 & 0.0495 & 99.62\% & 102.45\% & 83.33\% \\
LASSO(DD)-U & 0.0671 & 0.0488 & 100.97\% & 101.06\% & 90.00\% \\

OCMT(DD) & 0.0665 & 0.0490 & 100.03\% & 101.54\% & 73.33\% \\
OCMT(DD)-U & 0.0642 & 0.0475 & 96.70\% & 98.46\% & 96.67\% \\
\midrule
\multicolumn{6}{l}{\textbf{Horizon 8}} \\
AR & 0.0646 & 0.0467 & 100.00\% & 100.00\% & 96.67\% \\
ARX & 0.0712 & 0.0533 & 112.80\% & 114.27\% & 80.00\% \\
LASSO\_i & 0.0672 & 0.0500 & 103.68\% & 107.19\% & 83.33\% \\
LASSO(DD) & 0.0673 & 0.0493 & 103.79\% & 105.70\% & 90.00\% \\
LASSO(DD)-U & 0.0668 & 0.0491 & 103.00\% & 105.12\% & 80.00\% \\

OCMT(DD) & 0.0653 & 0.0475 & 100.85\% & 101.80\% & 80.00\% \\
OCMT(DD)-U & 0.0658 & 0.0483 & 101.37\% & 103.53\% & 90.00\% \\
\midrule
\bottomrule
\end{tabular}

\vspace{0.5cm}
{\footnotesize
\begin{minipage}{\linewidth}
\raggedright
\textit{Note:} LASSO(DD)-U and OCMT(DD)-U indicate dominant drivers sets selected without diagonal ratio restriction. Both RMSE and MAE ratios are compared against the AR model. MCS rate reports the fraction of countries in which the model belongs to the 10\% Model Confidence Set \citep{hansen2011model} computed on squared forecast errors using a moving-block bootstrap (block length 3, 2000 replications) and the range statistic based on the maximum absolute studentised pairwise mean loss differential across models. The procedure iteratively drops the model performs the worst in the comparison. A higher MCS rate implies the model is more competitive in terms of squared loss. 
\end{minipage}
}
\end{table}

%% file: Tables/forecast_presentation_table_gvar70_PC.tex
\begin{table}[htbp]
\centering
\caption{Forecast evaluation -- GVAR (Unobserved common factors controlled)}
\label{tab:forecast_gvar_PC}
\begin{tabular}{lccccc}
\toprule
Model & RMSE & MAE & RMSE ratio & MAE ratio & MCS rate \\
\midrule
\multicolumn{6}{l}{\textbf{Horizon 1}} \\
AR & 0.0612 & 0.0457 & 100.00\% & 100.00\% & 63.64\% \\
ARX & 0.0739 & 0.0592 & 127.26\% & 129.57\% & 63.64\% \\
LASSO\_i & 0.0642 & 0.0480 & 105.64\% & 105.17\% & 69.70\% \\
LASSO(DD) & 0.0602 & 0.0450 & 98.87\% & 98.48\% & 78.79\% \\
LASSO(DD)-U & 0.0561 & 0.0412 & 92.38\% & 90.27\% & 93.94\% \\
OCMT(DD) & 0.0594 & 0.0442 & 97.55\% & 96.91\% & 90.91\% \\
OCMT(DD)-U & 0.0559 & 0.0413 & 92.16\% & 90.44\% & 96.97\% \\
\midrule
\multicolumn{6}{l}{\textbf{Horizon 2}} \\
AR & 0.0604 & 0.0448 & 100.00\% & 100.00\% & 72.73\% \\
ARX & 0.0702 & 0.0540 & 118.14\% & 120.37\% & 63.64\% \\
LASSO\_i & 0.0602 & 0.0443 & 99.90\% & 98.84\% & 87.88\% \\
LASSO(DD) & 0.0577 & 0.0431 & 96.67\% & 96.04\% & 96.97\% \\
LASSO(DD)-U & 0.0564 & 0.0418 & 94.49\% & 93.12\% & 100.00\% \\
OCMT(DD) & 0.0597 & 0.0449 & 98.96\% & 100.13\% & 84.85\% \\
OCMT(DD)-U & 0.0563 & 0.0417 & 94.26\% & 93.00\% & 100.00\% \\
\midrule
\multicolumn{6}{l}{\textbf{Horizon 4}} \\
AR & 0.0591 & 0.0437 & 100.00\% & 100.00\% & 75.76\% \\
ARX & 0.0964 & 0.0774 & 261.54\% & 176.96\% & 51.52\% \\
LASSO\_i & 0.0611 & 0.0457 & 103.27\% & 104.40\% & 78.79\% \\
LASSO(DD) & 0.0587 & 0.0437 & 99.41\% & 100.00\% & 93.94\% \\
LASSO(DD)-U & 0.0582 & 0.0427 & 98.89\% & 97.64\% & 84.85\% \\

OCMT(DD) & 0.0597 & 0.0455 & 100.95\% & 103.96\% & 84.85\% \\
OCMT(DD)-U & 0.0576 & 0.0422 & 97.84\% & 96.60\% & 90.91\% \\
\midrule
\multicolumn{6}{l}{\textbf{Horizon 8}} \\
AR & 0.0577 & 0.0426 & 100.00\% & 100.00\% & 87.88\% \\
ARX & 0.0637 & 0.0499 & 114.82\% & 117.18\% & 81.82\% \\
LASSO\_i & 0.0592 & 0.0440 & 102.55\% & 103.22\% & 87.88\% \\
LASSO(DD) & 0.0601 & 0.0453 & 103.89\% & 106.42\% & 90.91\% \\
LASSO(DD)-U & 0.0585 & 0.0434 & 101.46\% & 101.89\% & 84.85\% \\

OCMT(DD) & 0.0588 & 0.0434 & 101.99\% & 101.97\% & 90.91\% \\
OCMT(DD)-U & 0.0581 & 0.0431 & 100.76\% & 101.24\% & 84.85\% \\
\midrule
\bottomrule
\end{tabular}

\vspace{0.5cm}
{\footnotesize
\begin{minipage}{\linewidth}
\raggedright
\textit{Note:} LASSO(DD)-U and OCMT(DD)-U indicate dominant drivers sets selected without diagonal ratio restriction. Both RMSE and MAE ratios are compared against the AR model. MCS rate reports the fraction of countries in which the model belongs to the 10\% Model Confidence Set \citep{hansen2011model} computed on squared forecast errors using a moving-block bootstrap (block length 3, 2000 replications) and the range statistic based on the maximum absolute studentised pairwise mean loss differential across models. The procedure iteratively drops the model performs the worst in the comparison. A higher MCS rate implies the model is more competitive in terms of squared loss. 
\end{minipage}
}
\end{table}